\documentclass[useAMS, usenatbib]{mn2e}
          
\usepackage[english]{babel}
\usepackage[dvips]{graphicx}
\usepackage{lscape}
\usepackage{amsmath}
\usepackage{amssymb}

\title[Stellar population gradients in Fornax S0 galaxies] 
  {Stellar population gradients in Fornax Cluster S0 galaxies: \\ 
   connecting bulge and disk evolution} 

\author[A.G. Bedregal, N. Cardiel, A. Arag\'on-Salamanca and M.R. Merrifield]
{A.G. Bedregal$^{1,2}$\thanks{E-mail: bedregal@astrax.fis.ucm.es, bedregal@astro.umn.edu}, 
 N. Cardiel$^{1}$, A. Arag\'on-Salamanca$^{2}$ and M.R. Merrifield$^{2}$\\
  $^{1}$Departamento de Astrof\'{\i}sica, Facultad de Ciencias
  F\'isicas, Universidad Complutense de Madrid, 28040 Madrid, Spain\\
  $^{2}$School of Physics and Astronomy, University of Nottingham,
  University Park, Nottingham, NG7 2RD, UK}

\begin{document}

\date{Accepted ***. Received ***; in original form ***}

\pagerange{\pageref{firstpage}--\pageref{lastpage}} \pubyear{2010}

\maketitle

\label{firstpage}

\begin{abstract}
  We present absorption-line index gradients for a sample of S0
  galaxies in the Fornax Cluster.  The sample has been selected to
  span a wide range in galaxy mass, and the deep VLT-FORS2
  spectroscopy allows us to explore the stellar populations all the
  way to the outer disk-dominated regions of these galaxies.  We find
  that globally, in both bulges and disks, star formation ceased
  earliest in the most massive systems, as a further manifestation of
  downsizing.  However, within many galaxies, we find an age
  gradient which indicates that star formation ended first in the
  outermost regions.  Metallicity gradients, when detected, are always
  negative such that the galaxy centres are more metal-rich.  This
  finding fits with a picture in which star formation continued in the
  central regions, with enriched material, after it had stopped in the
  outskirts.  Age and metallicity gradients are correlated, suggesting
  that large differences in star formation history between the inner
  and outer parts of S0 galaxies yield large differences in their
  chemical enrichment.  In agreement with previous results, we
  conclude that the radial variations in the stellar populations of S0
  galaxies are compatible with the hypothesis that these galaxies are
  the descendants of spiral galaxies whose star formation has
  ceased. With the addition of radial gradient information, we are
  able to show that this shutdown of star formation occurred from the
  outside inward, with the later star formation in the central regions
  offering a plausible mechanism for enhancing the bulge light in
  these systems, as the transformation to more bulge-dominated S0
  galaxies requires.
\end{abstract}


\section{Introduction}  

In order to better understand the formation and evolution of S0
galaxies, we have undertaken a detailed study of these
systems in the Fornax Cluster, combining deep optical long-slit
spectroscopy obtained with the FORS instrument on the VLT with archival
optical and near-infrared imaging.  The sample and major-axis
kinematics of these galaxies are presented in Bedregal et al.\,(2006a,
hereafter Paper\,I).  In Bedregal et al.\,(2006b, hereafter
Paper\,II), we used those data to study the Tully--Fisher relation
(Tully \& Fisher 1977) for these systems as part of a larger
compilation of 60 local S0s, and concluded that their disk kinematics
were consistent with what would be expected if these systems were
simply faded S0 galaxies.  In Bedregal et al.\,(2008, hereafter
Paper\,III) we studied the central stellar population properties of
these galaxies, and found evidence that the bulge-dominated regions of
these galaxies had star-formation properties that varied
systematically with the mass of their host galaxies.  In this paper,
we seek to tie together these bulge and disk properties by studying
the gradients in common absorption-line indices.  By obtaining such
gradients out as far as the region where the disk dominates the light,
and interpreting them in terms of simple stellar population models, we
seek an integrated understanding of the formation and evolution of
both of the major components of S0 galaxies.

It is already well known that elliptical galaxies and S0 systems show
variations in the properties of their stellar populations with radius.
Early studies using broad-band colours established that the nuclear
regions are usually redder than the outskirts, suggesting radial
variations in their stellar population properties (Kormendy \&
Djorgovski 1989, and references therein).  The first study of
gradients in spectral absorption features was performed by McClure
(1969): in a sample of 7 galaxies, he found that an index measuring
the central CN$\lambda$4216 band strength was stronger than at larger
radii, which he interpreted as the result of metallicity variations
between the different regions. Subsequent spectroscopic studies, many
based on the Lick/IDS line strength indices (Worthey et al.\,1994),
have explored absorption feature gradients using a broad range of line
indices (e.g.\,Spinrad et al.\,1972; Welch \& Forrester 1972; Joly \&
Andrillat 1973; Cohen 1979; Couture \& Hardy 1988; Peletier 1989;
Gorgas et al.\,1990; Bender \& Surma 1992; Davies et al.\,1993;
Carollo et al.\,1993; Gonz\'alez 1993; Fisher et al.\,1995; Gorgas et
al.\,1997; Mehlert et al.\,2003; S\'anchez-Bl\'azquez et al.\,2006,
2007). Overall, these studies indicate the presence of strong
gradients in CN features (at 3883 and 4216$\,$\AA), weaker gradients
in the Mg-triplet (at $\approx 5200\,$\AA), g-band, Na D lines, Ca
H\&K features and some FeI sensitive indices, and no gradients in the
hydrogen Balmer series (e.g.\,H$\beta$, H$\gamma$), MgH, TiO, CaI and
the infrared Ca triplet.  Many of these studies suggest that the
existence of gradients in metallic features is produced by radial
variations of metallicity, decreasing outward (e.g.\,Cohen 1979;
Davies et al.\,1993; Kobayashi \& Arimoto 1999; Mehlert et al.\,2003).
Other works, however, suggest that there is also a gradient in the
stellar population age, the central regions being younger than the
outer parts (e.g.\,Gorgas et al.\,1990; Munn 1992; Gonz\'alez 1993;
Gonz\'alez \& Gorgas 1996).

These absorption-line index gradients provide clues to the mechanisms
responsible of galaxy formation. Chemodynamical models of scenarios
such as monolithic collapse, merger-driven evolution and other secular
processes predict different gradients.  Although quantitative modeling
of these effects is still in its infancy, clear qualitative indicators
are becoming established.  For example, monolithic collapse models
(e.g.\,Eggen, Lynden-Bell \& Sandage 1962; Larson 1974a; Carlberg
1984; Arimoto \& Yoshii 1987) generically predict the formation of
strong gradients for both age- and metallicity-sensitive indices,
while, in comparison, a formation history driven by mergers will weaken any such
gradients significantly (e.g.\,White 1980; Bekki \& Shioya 1999;
Barnes \& Hernquist 1991; Mihos \& Hernquist 1994).

Clues to the more detailed star formation history of a galaxy can be
obtained by comparing individual chemical elements.  The different
timescales associated with the supernova producing $\alpha$-elements
(such as Mg) and Fe mean that two galaxies with the same mean stellar
age but different histories will display different gradients in these
elements.  For example, a galaxy in which the stars were produced by a
short starburst in the distant past and one where the star formation
was over a more extended period may display similar constant Fe index
values with radius, but the former will show a decreasing gradient in
Mg line strengths while the latter will display a flatter Mg index
profile (e.g.\,Thomas 1999).

Thus, absorption-line index gradients offer a valuable tool in the
study of galaxy formation and evolution.  Although we are not yet in a
position to evaluate such gradients in an unequivocal quantitative
manner, their basic interpretation is fairly clear.  Even without
detailed modeling, their differential use is robust in that systematic
differences between galaxies of different types are generally
unambiguous.  In addition, significant progress is being made with the
quantitative analysis of such data, through the development of more
sophisticated stellar population models.  Hence the philosophy of this
work is initially to make use just of the directly-observable
line-strength gradients of the Fornax S0s to seek qualitative insight
into any systematic variations in their evolutionary histories, and
then to go on to use the new stellar population models of Vazdekis et
al.\,(2010) to interpret such results in a more quantitative manner.

The remainder of the paper is as laid out follows. In Section\,2, we
present the line index gradients derived from the deep long-slit
spectroscopy of 9 Fornax Cluster S0 galaxies, with the associated
detailed plots and values for individual galaxies provided in
Appendix\,A.  Section\,3 describes the use of the new stellar
population models to obtain ages and chemical abundances (and their
gradients) from the observed line indices, with the detailed data
given in Appendix\,B.  In Section\,4, we discuss the implications of
the results on the formation history of S0 galaxies, while Section\,5
summarizes these results.

\section{Line strength index gradients}\label{sec:GradInd}

In this section, the index gradients measured for ten common
absorption lines are presented.  These are the same indices derived
for the centres of these galaxies in Paper\,III, and the details of
how they and their associated uncertainties were measured are
presented in that paper. A description 
of how the spectra were binned in 
the spatial direction is also presented in Paper\,III, Sec.\,2. The binning 
was designed to yield a minimum signal-to-noise ratio in each spatial bin 
so that reliable spectral indices could be measured. See Paper\,III for details.
Note that for consistency all indices are
calculated on a magnitude scale. 

In Appendix\,A, Figs.\,\ref{fig:IndGradN1316} to \ref{fig:IndGradE002}
show line index measurements along the photometric major axes of the
sample galaxies.  The radial profiles have been folded around their
galaxies' centres to give a sense of any systematic variations. The
radii have been scaled to each galaxy's half-light radius (hereafter, $\rm R_{HALF}$) as
calculated from infrared photometry (in the $\rm K_{\rm s}$-band; see
Paper\,II). The straight lines show linear fits of these logarithmic
parameters, with those points lying at radii small enough to be
affected by seeing excluded from the fit.  The resulting values of the
index gradients and their uncertainties are presented in
Table\,\ref{tab:GradInd}.

\begin{figure*}
\begin{center}
\includegraphics[scale=1.10]{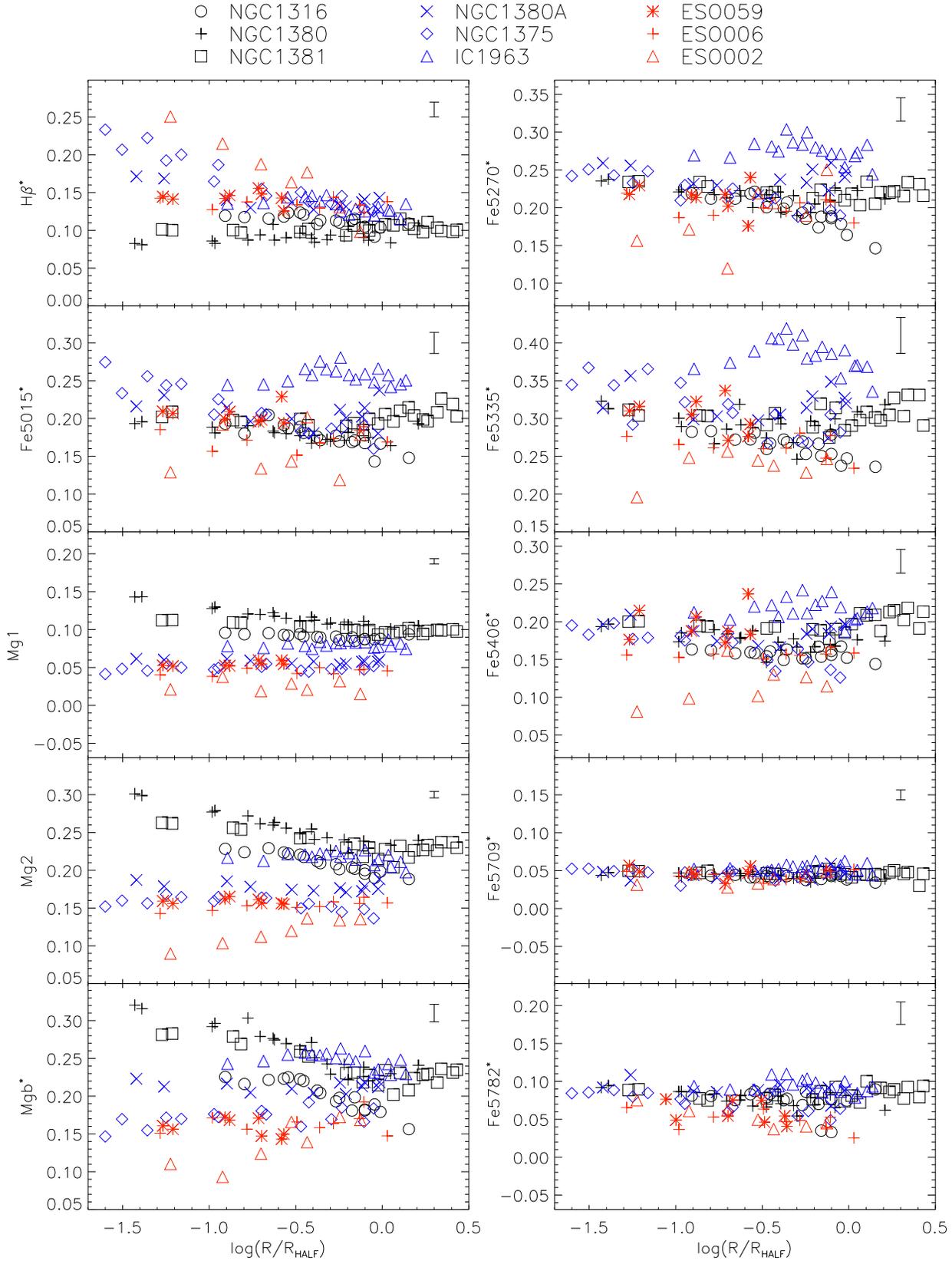}
\end{center}
\caption{\label{fig:rad_v_Indices_allS0s} Measurements of 10
  line-strength indices as a function of radius for the entire galaxy
  sample.  The key to the symbol representing each galaxy is shown at the
  top of the figure. The color code split the galaxy sample by
  luminosity into a bright (black), intermediate (blue) and faint
  (red) sample.  For clarity, the error bars on each point, shown in
  the individual profiles of Figs.\,\ref{fig:IndGradN1316} --
  \ref{fig:IndGradE002}, are not shown. Instead, in the top-right corner of 
   each panel we present the median $\rm \pm 1\,\sigma$ uncertainty of the measured indices.}
\end{figure*}

\begin{figure*}
\begin{center}
\includegraphics[scale=0.72, angle=90]{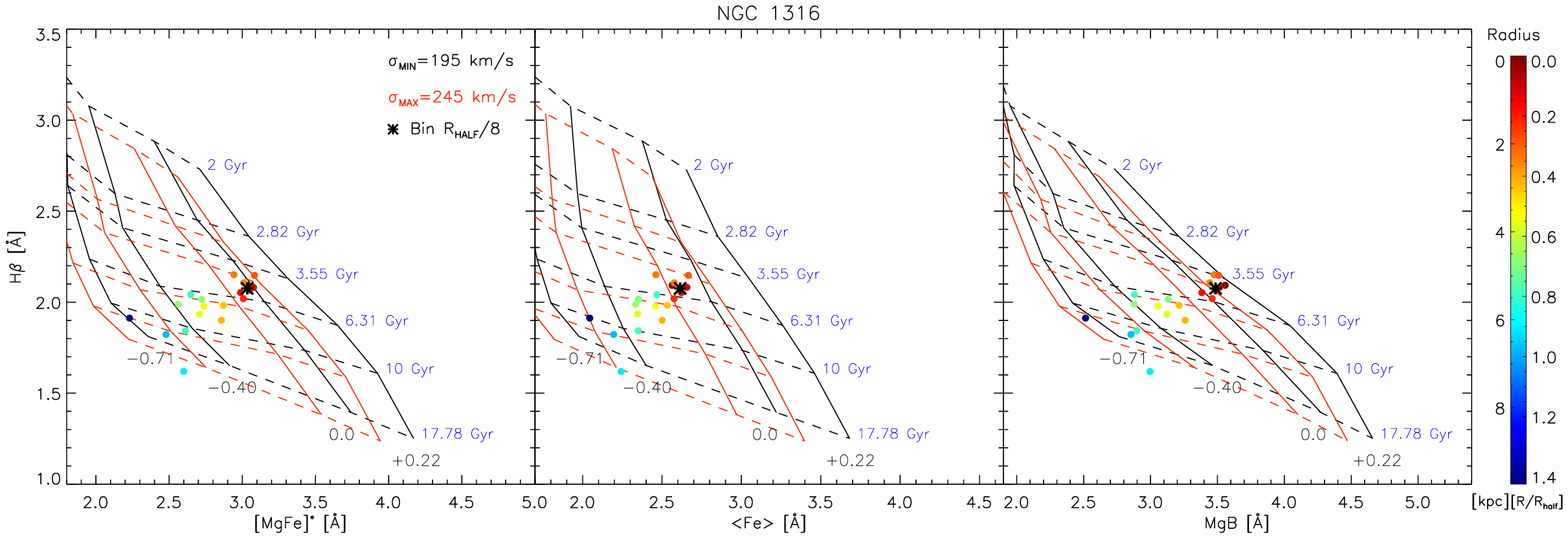}
\end{center}
\caption{\label{fig:n1316_3MILES}  Example of NGC\,1316 radial data with Vazdekis et al.\,(2010) SSP model grids. H$\beta$ is used as the age-sensitive index, while [MgFe]$^{\prime}$ (left), $\rm \langle Fe\rangle$ (center) and Mg$b$ (right) are used as [M/H]-sensitive indicators. Color code on galaxy data represent the radial distance of each data point. The black asterisks represent the central measurements (within $\rm R_{HALF}/8$). For clarity, only two model grids are shown, those with the higher (red) and lower (black) galaxy velocity dispersion.}
\end{figure*}

The line index gradients sample different fractions of the bulge- and disk-dominated regions of these galaxies. In Paper\,I we calculated a limiting radius ($\rm R_{LIM}$, dot-dashed blue lines in Figs. \,\ref{fig:IndGradN1316} to \ref{fig:IndGradE002}) from which the light profile is dominated by the disk component. Therefore, with $\rm R_{LIM}$ we can assess the contribution of each component to the measured gradients. The relative bulge and disk contribution to the gradients is rather heterogeneous among our galaxy sample. For three galaxies the number of radial bins dominated by disk light is less than 30\% (NGC\,1380, 1381 and 1375). For two other systems the disk contribution is between 70 and 90\% (NGC\,1380A and IC\,1963), while other two galaxies show gradients totally dominated by the disk light (ESO\,358-G006 and ESO\,359-G002). Finally, NGC\,1316 and ESO\,358-G059 have no information in this regard as $\rm R_{LIM}$ cannot be estimated in those systems where the disk never dominates the luminosity profiles.

In order to summarize this mass of information and search for
systematic variations between Fornax Cluster S0s,
Fig.\,\ref{fig:rad_v_Indices_allS0s} shows over-plotted the radial
variation in all the line indices for all the sample galaxies.  The
galaxies have been divided and colour-coded by luminosity:
\begin{itemize}
\item {\it Bright S0s} ($ -19.0\le  M_{\rm B} \le -20.6$): NGC\,1316, 1380
  and 1381, shown as black symbols.
\item {\it Intermediate S0s} ($ -18.1\le  M_{\rm B}  \le -18.5$):
  NGC\,1380A, 1375 and IC\,1963, shown as blue symbols.
\item {\it Faint S0s} ($-17.3\le  M_{\rm B}  \le -17.5$):
  ESO\,358-G059, ESO\,358-G006 and ESO\,359-G002, shown as red symbols.
\end{itemize}
When interpreting the results we need to take into consideration 
that NGC\,1316 (in the bright S0 category) shows clear evidence of
having suffered a recent merger (Schweizer 1980; Goudfrooij et al.\
2001a,b; Paper~I), so may not be representative of the class.  As is
clear from Fig.~\ref{fig:rad_v_Indices_allS0s}, there are strong
variations in behaviour both between different line indices and
between different galaxies, but some systematic trends are immediately
apparent.  We therefore seek to explore those systematic effects more
fully.

One initial point to note is that most of the stronger gradients measured 
are driven by the inner parts of bulge-dominated regions (e.g.~Mg indices in 
NGC\,1380, 1381, H$\beta$ in NGC\,1375). Generally, the radial profiles become fairly flat even
before reaching $\rm 1\,R_{HALF}$, immediately suggesting a
scenario in which bulge and disk components may have undergone rather
different mass assembly and star formation histories.  However, the
radial behaviour of the indices present strong variations from galaxy
to galaxy, suggesting a degree of variation in any such history.  In
addition, the different line indices behave in rather different ways:
all panels in Fig.\,\ref{fig:rad_v_Indices_allS0s} have vertical widths
of 0.3\,mag to allow a fair visual comparison between radial index
profiles, and so it is apparent that the H$\beta$ and Mg indices
display a wide variety of radial gradients, while the Fe tracers have
much flatter profiles, but with a range of absolute values.

Perhaps the most interesting phenomenon is that the H$\beta$ index,
while varying widely from galaxy to galaxy at small radii, seems to
converge to a common low value at large radii.  The lower-mass systems
have systematically steeper gradients in this index, and
correspondingly higher central values, while the higher-mass galaxies
have similar low values of H$\beta$ at all radii.  The simplest
interpretation of this effect is that the lower-mass galaxies have
undergone their final episode of star formation more recently, and that
this star formation occurred primarily at small radii. It is possible that the
final star formation event in the higher-mass galaxies was similarly
centrally concentrated, but occurred sufficiently long ago that any
gradient in H$\beta$ has now disappeared.  We also need to be a little
careful because variations in the H$\beta$ line index arise from both
metallicity and age effects.  We therefore now turn to interpreting
the physical significance of the H$\beta$ profile and those of the other line indices.

\section{Single Stellar Population Parameters and their Gradients}\label{sec:data:ind}

Given the complex dependency of the observed absorption line indices
on stellar population properties such as star-formation history and
the chemical abundances of different elements, we rely on somewhat
simplified models to translate the observations into 
parameters that have at least a crude physical interpretation.  In
particular, by fitting these indices to models of a single stellar
population (SSP) with all stars having a common age and chemical
abundances, we can derive a luminosity-weighted measure of the age,
metallicity and $\alpha$-abundance of the observed population.  

In this section we will present the steps required to estimate these
parameters using the new SSP models of Vazdekis et al.\,(2010) based
on the {\tt MILES} stellar library (S\'anchez-Bl\'azquez et al.\,2006;
Cenarro et al.\,2007).  This library, with its $\sim$1000 stellar
spectra, provides very good coverage of the stellar atmosphere
parameter space (gravity, temperature and metallicity) at a much
higher spectral resolution than the original Lick/IDS stellar library:
with an instrumental dispersion of $\approx 56\,{\rm
  km}\,{\rm s}^{-1}$, it is well matched to the data analyzed here,
allowing us, in many cases, to use the full observed spectral resolution when
comparing data and models. Indeed, the velocity dispersion of the
galaxy data implies that in many situations we will be
degrading the SSP models to the resolution of the galaxy spectra, and
not the other way round, thus minimizing the loss of information.
This optimal use of the data is particularly important in the fainter
low-mass galaxies and in the disk-dominated outer regions of all the
systems, as these regions share the observationally-challenging
properties of low surface brightness and low intrinsic velocity
dispersion.  One further benefit of the {\tt MILES} data is that they
are flux calibrated, avoiding the complex and somewhat uncertain
corrections involved in placing the absorption-line indices in the
Lick/IDS library system. Note that the Mg$_1$ and Mg$_2$ indices
presented in Paper\,III were corrected to the Lick/IDS system, but the
ones used here are not.

The Vazdekis et al.\,(2010) models make full use of the {\tt MILES}
library and incorporate the latest developments in stellar evolution
and SSP model techniques.  In addition, they offer a
publicly-available easily-usable web-based code
(http://miles.iac.es/), which means that the analysis here can readily
be reproduced and compared to the results from other data sets, thus
making it the ideal choice for this analysis.

\subsection{Stellar Population Parameters}\label{subsec:StelPopCalc}
We estimate SSP ages and metallicities ([M/H]) using Vazdekis et
al.\,(2010) model predictions and assuming a Kroupa-Revised initial
mass function (IMF; Kroupa et al.\,2003).  The model spectra,
corresponding to different combinations of age and [M/H], were
matched to the spectral resolution for each galaxy and
radial position by convolving them with a Gaussian of dispersion
\begin{equation}
\sigma = \sqrt{\sigma_{\rm gal}^{2} - \sigma_{\rm lib}^{2}},\label{eq:broad}
\end{equation}
where $\sigma_{\rm gal}$ is the velocity dispersion of the galaxy at
the corresponding radius (see Paper\,I), and $\sigma_{\rm lib}$ is the
resolution from the stellar library.  Since $\sigma_{\rm lib}$ has a
small, but significant, wavelength dependence, the model spectra were
degraded separately for each of the absorption lines used as the main
diagnostics (H$\beta$, Mg$b$, Fe5270 and Fe5335) using the value
appropriate for the central wavelength of each feature. In the cases
where $\sigma_{\rm gal} < \sigma_{\rm lib}$ (NGC\,1380A, IC\,1963,
ESO\,358-G059, ESO\,358-G006 and the outskirts of NGC\,1381, NGC\,1375
and ESO\,359-G002), the galaxy spectra were degraded to reach the
library's resolution by convolving them with a Gaussian of dispersion
\begin{equation}
\sigma = \sqrt{\sigma_{\rm lib}^{2} - \sigma_{\rm inst}^{2} - \sigma_{\rm gal}^{2}},\label{eq:broad2}
\end{equation}
where $\sigma_{\rm inst}$ is the FORS2 instrumental dispersion of
$\sim$30\,km$\,$s$^{-1}$.  

Once the model and galaxy spectra were matched in resolution, their
line strength indices were measured.  To estimate ages and [M/H],
model grids for each individual galaxy data point were created as
illustrated by the example shown in Fig.\,\ref{fig:n1316_3MILES}.  In
this figure we show the radial data of NGC\,1316 with Vazdekis et
al.\,(2010) SSP model grids for 3 different metallicity tracers. The
color code of the galaxy data represents the corresponding
galactocentric distance.  Note that we only show the grids for the
models with the largest and smallest velocity dispersions to
illustrate the amplitude of this effect; for the actual fitting, each
data point was matched to its own grid of models.  In the final
modeling, the parameters were calculated using H$\beta$ as the main
age-sensitive index.  Given the range of $\alpha$-element abundances
hinted at by the analysis of Section\,\ref{sec:GradInd}, we adopt the
combined [MgFe]$^{\prime}$ index as the primary measure of [M/H], due
to its negligible sensitivity to $\alpha$-element abundance
(Gonz\'alez 1993; Thomas, Maraston \& Bender 2003).  The best estimate
SSP values of age and {M/H] were then obtained from this grid using a
bivariate polynomial interpolation between the calculated grid
values.  Finally, any variation with radius in the relative $\alpha$-element
abundances is traced using the  Mg$b$/$\rm \langle Fe\rangle$ ratio
(see Paper\,III and Fig.\,4 of Thomas, Maraston \& Bender 2003).  
The resulting profiles for age, [M/H] and $\alpha$-enhancement 
with radius for all of the sample galaxies are presented
in Appendix\,B, Figs.\,\ref{fig:PGradN1316} to \ref{fig:PGradE002}.

\subsection{Uncertainties}\label{subsec:AZError}

\begin{figure}
\begin{center}
\includegraphics[scale=.7]{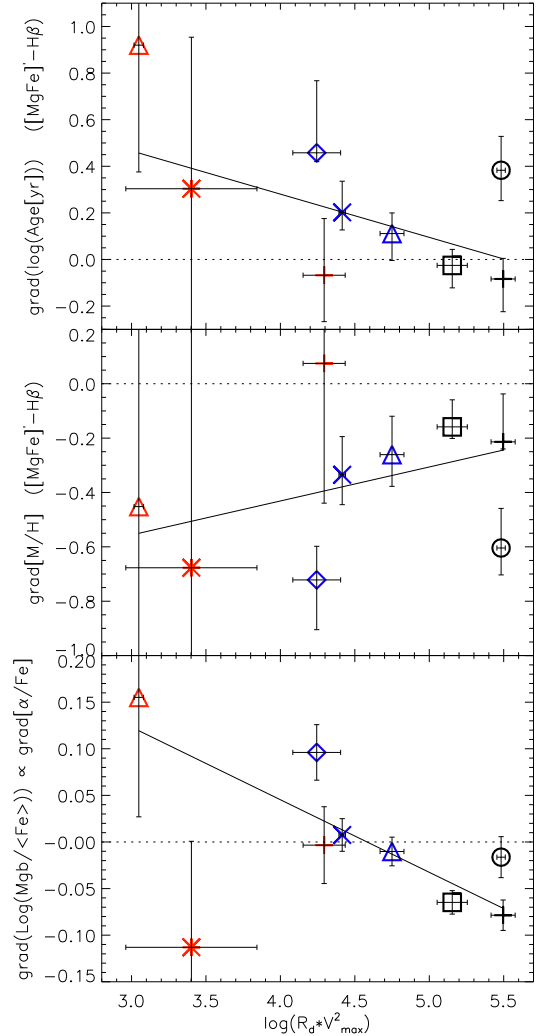}
\end{center}
\caption{\label{fig:Grad_vs_Mass_allS0s}  Gradients in Age, [M/H] and [$\alpha$/Fe] tracer versus 
$\log (R_{\rm d} \cdot v_{\rm max}^2)$, a proxy for dynamical mass. 
(see Paper\,III). Continuous lines correspond to best linear fits. 
Dashed lines represent flat gradients. Symbols as in Fig.\,\ref{fig:rad_v_Indices_allS0s}.}
\end{figure}

When interpreting any correlations apparent in these results, it is
important to realize that the nature of the modeling process means
that the estimated parameters are not independent.  We must therefore
take some care in estimating the associated uncertainties.  The
errors in individual data points were calculated by Monte Carlo simulation, generating pairs
of values around the measured H$\beta$ and [MgFe]$^{\prime}$ indices,
with a dispersion matched to the observational errors on these
quantities, placing them on the model grid, and then extracting the
age and  [M/H] parameters.  The 68th percentile of the resulting
age and [M/H] distributions then provides a realistic quasi-elliptical
1\,$\sigma$ error contour around the derived Age-[M/H] value. The
associated (often asymmetric) errorbars can then be derived by
projecting the error contours onto the age and [M/H] axes.

Similarly, the uncertainties in the Mg$b$/$\rm \langle Fe\rangle$ 
ratio (our $\alpha$-abundance tracer) were estimated by Monte Carlo simulations, 
producing symmetric error bars in this case.

\subsection{Gradients}\label{subsec:TheGradients}

Figs.\,\ref{fig:PGradN1316}-\ref{fig:PGradE002} in Appendix\,B show
the radial variation of the derived ages, [M/H] and
$\alpha$-enhancement indicators for all the galaxies, together with their
uncertainties.  Linear fits to those gradients are plotted as solid
lines.  The slopes of these fits (hereafter called ``gradients'') are
given in Table~\ref{tab:GradAZA}.

The uncertainties in age and [M/H] gradients were
calculated using the Monte Carlo simulations described above for every
H$\beta$-[MgFe]$^{\prime}$ pair, translated into simulated age-[M/H]
pairs.  From these values we created a large number of realizations of
the radial distributions of these two parameters for every galaxy,
each of which was fitted with a linear function to give an estimate
of the uncertainty in each gradient (and its correlation with other
parameters).  The quoted errors in these two parameters are asymmetric 
as a consequence of using individual age and [M/H] predictions having 
asymmetric uncertainties (see Sec.\,\ref{subsec:AZError}). Errors in the
$\alpha$-enhancement gradients, also calculated via Monte Carlo simulations, 
are  symmetric since no models are involved in their determination.

\section{Results}\label{sec:Results}

There are almost no previous studies of the stellar population 
gradients in S0s where these galaxies are treated as a
separate class from the ellipticals (hereafter, Es). The
work of Fisher, Franx \& Illingworth (1996) is one
of the few exceptions. For a sample of 9 bright edge-on S0s they
found that their bulge and disk metallicity gradients differ
substantially, being steep and decreasing outwards in the
bulge region while becoming flat in the stellar disk. They also
found that the central Mg and Fe indices correlate with central
velocity dispersion (hereafter, $\rm \sigma_0$) as in luminous Es.
Hence, they conclude that S0 bulges are probably the result of a dissipative
collapse process (such as gas-rich  mergers) at very early times.

For a combined sample of 35 E and S0 galaxies in the Coma cluster, 
Mehlert et al.\,(2003) calculate gradients reaching $\rm \approx 1\,R_{HALF}$. 
They found that Es have on average slightly higher velocity dispersions, 
lower H$\beta$ and higher metallic line-strengths than S0s. 
The latter form two families, one comparable to the Es and a second one 
with much younger stellar populations ($\rm \sim 2\,Gyr$), and weaker metallic lines. 
This result was later confirmed by Kuntschner et al.\,(2006)  
within the {\tt SAURON} collaboration (Bacon et al.\,2001).
These authors studied a sample of 48 local Es and S0s  
and identified the young, low-$\rm \sigma_0$ S0s as the main drivers 
of the observed trends between gradients and central line indices.

In the light of these results we now analyse 
the gradients found in our Fornax S0 sample.
The first point apparent from Table~\ref{tab:GradAZA} is that
virtually all the age gradients are positive or flat, indicating that
the star formation ended first in the disk-dominated outer regions of
the galaxies.  This result confirms the qualitative indication that we
found from the H$\beta$ gradients shown in
Fig.\,\ref{fig:rad_v_Indices_allS0s}.  Indications of this result were
also found in Paper\,II, and we argued there that
centrally-concentrated star formation at later times, enhancing the
bulge-to-disk ratio, is probably an integral stage in the
morphological transformation of a spiral into a S0 galaxy (Christlein
\& Zabludoff 2004; Shioya, Bekki \& Couch 2004).  Other observational
evidence in favour of this scenario has been presented in the past
(Moss \& Whittle 2000; Mehlert et al. 2003; Poggianti et al. 2001;
Bamford, Milvang-Jensen, \& Arag\'on-Salamanca 2007).  What is
still unclear is whether this late central star formation represents
an enhancement over the ``quiescent spiral'' star formation rate
(perhaps fueled by interaction-induced inward gas flows) or is just
the last remaining centrally-concentrated star-formation gasp.  The
second hypothesis is probably favoured by some recent results (e.g.,
Wolf et al.\ 2009) but the evidence is still inconclusive.

Figure\,\ref{fig:rad_v_Indices_allS0s} also suggested that the the
cessation of star formation happened at earlier times in the brightest
galaxies, considerably reducing the H$\beta$ and age gradients. If
that trend is a reflection of an underlying galaxy mass dependency we
would expect an anti-correlation between age gradient and galaxy mass.
To test for this effect, Fig.\,\ref{fig:Grad_vs_Mass_allS0s} shows the
various gradients plotted against $R_d\times v_{\rm max}^2$, which
provides a measure of each galaxy's mass (see Paper\,III).  There does
seem to be a trend between age gradient and mass in the sense
described (a Spearman rank test gives a 
probability of correlation between 95 and 97.5\%), although NGC\,1316, 
whose central regions have probably
been rejuvenated by a recent merger, seems to deviate somewhat. 
Possible correlations between the gradients in [M/H] or
$\alpha$-enhancement and galaxy mass are either weaker or not detected 
(Spearman rank tests give $<$90\% and 90-95\% probabilities, respectively).  
Previous studies (Mehlert et al.\ 2003; Rawle et al.\ 2010) found no 
correlations between these parameters in early-type galaxies (although 
see Spolaor et al.\,2009 and Kuntschner et al.\,2010 at low masses). In any case, some
care is necessary in not over-interpreting these trends (or lack
thereof) given the limitations imposed by the sample size and
uncertainties.

Close inspection of Figs.\,\ref{fig:PGradN1316} to \ref{fig:PGradE002}
also reveals that the generally old age that we had ascribed to the
disks in Section~\ref{sec:GradInd} is not the whole story, in that the
age in the outer disk-dominated regions is $\gtrsim10\,$Gyr for the
brightest and most massive galaxies, while it is only
$\simeq4$-$5\,$Gyr for the less massive ones.  This would seem to be a
new manifestation of ``downsizing'': star formation stopped first in
the most massive galaxies not only in their centres [as found before
in other early-type galaxy samples; see Smith at al.\,(2009) and
references therein] but also in their oldest outermost regions.

In Paper\,III we reported a tentative correlation between the
$\alpha$-element overabundance and age in the central regions of these
galaxies. This correlation is confirmed here with the improved models.
A Spearman rank test gives a probability of correlation between 95 and 97.5\%
(see Fig.\,\ref{fig:GradVSGrad_CentVSCent_allS0s} top-central panel). The correlation
can be explained by considering a simple model where all galaxies
started forming stars at approximately the same time, formed stars for
different time intervals $\Delta t$, and then ceased their star
formation $t\,$Gyr ago.  Since the SSP ages are correlated with $t$,
while the $\alpha$-element overabundance is anti-correlated with
$\Delta t$, it follows that the ages must be correlated with the
$\alpha$-element overabundance, as observed (see Paper\,III and
Arag\'on-Salamanca 2008 for details).
Fig.\,\ref{fig:GradVSGrad_CentVSCent_allS0s} also confirms Paper\,III
finding that there is no correlation between central ages and central
[M/H], nor between central [M/H] and central
$\alpha$-enhancements.

\begin{figure*}
\begin{center}
\includegraphics[scale=.7]{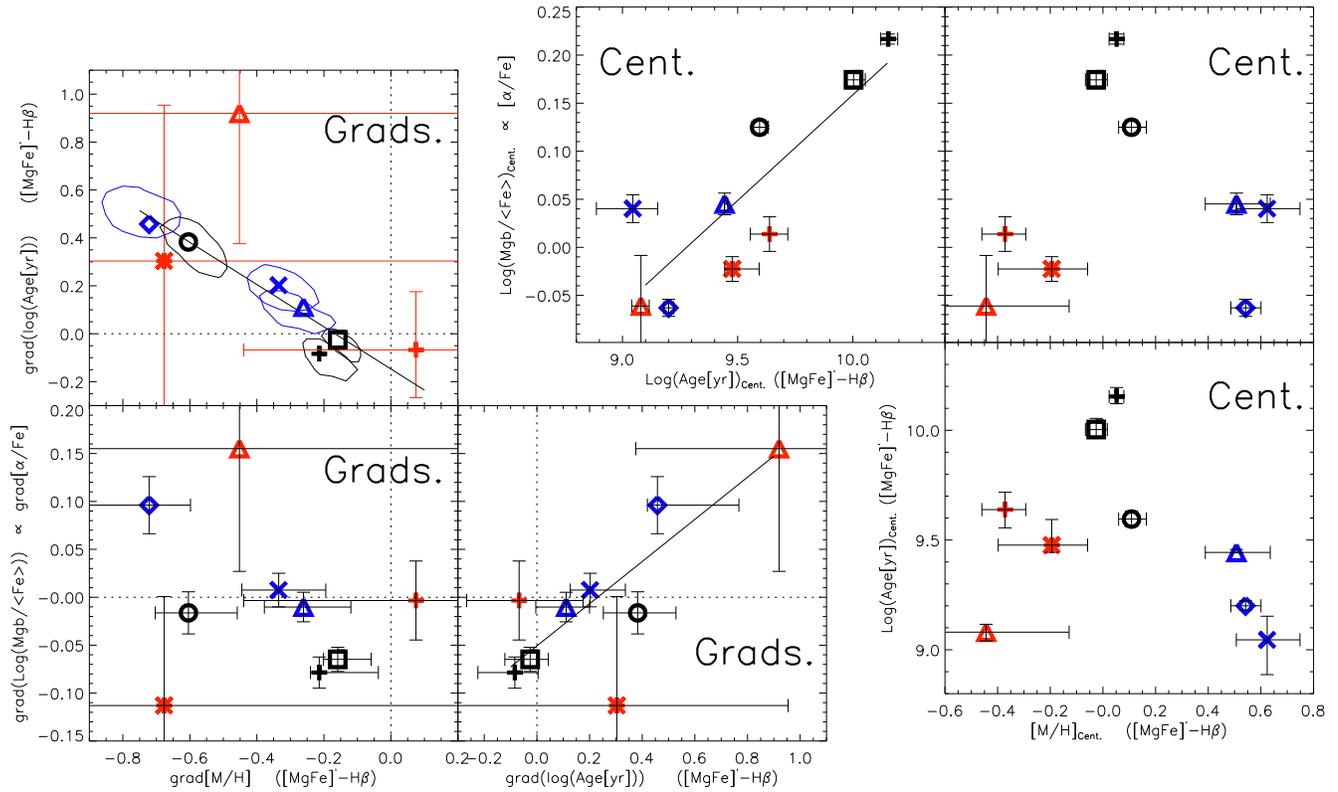}
\end{center}
\caption{\label{fig:GradVSGrad_CentVSCent_allS0s}  Gradients in Age, [M/H] and $\alpha$-enhancement tracer versus each other and Central Age, [M/H] and $\alpha$-enhancement tracer versus each other. Symbols as in Fig.\,\ref{fig:rad_v_Indices_allS0s}. Continuous lines correspond to the best linear fits. Dashed lines correspond to flat gradients. In Grad(Age) vs.\,Grad[M/H] (top-left panel), errors as 1\,$\sigma$ contours have been drown for some galaxies.}
\end{figure*}

When detected, the [M/H] gradients are always negative, implying that
the stars in the inner regions of the galaxies are on average more
metal-enriched than those in the outskirts. This finding fits in with
the growing picture that star formation continued, with enriched
material, in the central regions after it had stopped in the
outskirts,

Gradients in the $\alpha$-element enhancement tracer 
are usually flat within the uncertainties except for the two old 
and bright S0s (NGC\,1380, 1381). These galaxies show clear negative gradients (Table~\ref{tab:GradAZA}). 
In the bottom-central panel of Fig.\,\ref{fig:GradVSGrad_CentVSCent_allS0s} we see a trend between 
Grad(Age) and Grad($\alpha$). This trend is consistent with the similar correlation found 
for the central values (a Spearman rank test gives a 90-95\% probability of correlation).

Finally, the top-left panel of Fig.\,\ref{fig:GradVSGrad_CentVSCent_allS0s} clearly shows
that gradients in age and [M/H] are correlated for
the six brightest galaxies (black and blue points). Nothing can be
said for the faintest three, since no gradients were detected in them
within the large errors (see Table~\ref{tab:GradAZA}). A Spearman rank tests gives a 99.5\% probability of correlation for all 9 S0s.

It could be argued that, at least in part, this trend is driven by 
correlated errors as the 1\,$\sigma$ error ellipses in 
Fig.\,\ref{fig:GradVSGrad_CentVSCent_allS0s} are clearly aligned with this trend. 
We have tested whether this correlation can be totally explained by the observed 
uncertainties using $10^4$ Monte Carlo simulations which consider the observed data points and 
their error distributions. Using a Kolmogorov-Smirnov test with $\alpha=0.05$ confidence level
we find that 93\% of the simulations reject the 
null hypothesis that our original sample and each simulated one were drawn 
from the same distribution. This indicates that, 
although there is a sizable correlation  in the errors, there is a real  
correlation between  Grad[M/H] and Grad(Age).

The best linear fit to the Grad(Age) versus Grad[M/H] trend produces a slope of $\rm -1.14 \pm 0.30$. 
This result is comparable with findings of S\'anchez-Bl\'azquez et al.\,(2007; $\rm -1.33 \pm 0.37$). However, other works on early-type galaxy samples give somewhat different results, like Mehlert et al.\,(2003; $\rm -0.95 \pm 0.11$) and Rowle et al.\,(2010; $\rm -0.69 \pm 0.10$. The concordance with S\'anchez-Bl\'azquez et al.~result might be attributed to the fact that their sample, like ours, has an important fraction of disky, low-$\sigma_0$ galaxies.  Although difficult to interpret in detail, this correlation indicates that large differences in star formation history between the inner and outer
parts of galaxies yield large differences in chemical enrichment,
which in retrospect is perhaps not so surprising.

\section{Summary}

Using deep VLT-FORS2 spectroscopy we have derived absorption-line
index gradients for a sample of S0 galaxies in the Fornax Cluster
spanning a wide range in galaxy mass. Combining these observations
with the latest spectral synthesis models (Vazdekis et al. 2010),
we have studied the properties of the stellar populations in these galaxies
as a function of galactocentric distance, reaching the outer
disk-dominated regions.  

Our main conclusions are:
\begin{enumerate}
\item The age gradients are positive or flat, indicating that the star
  formation ended first in the disk-dominated outer regions of the
  galaxies.  This centrally-concentrated star formation at later times
  could serve to enhances the bulge-to-disk ratios of these systems,
  helping the morphological transformation of spirals into S0s.

\item The cessation of star formation in the central regions
  happened at earlier times in the brightest galaxies, considerably
  reducing their H$\beta$ and age gradients.

\item The age in the outer disk-dominated regions is $\gtrsim10\,$Gyr
  for the brightest and most massive galaxies and $\simeq4$-$5\,$Gyr
  for the least massive ones. Thus, star formation stopped first in the
  most massive galaxies not only in their centres but also in their
  oldest outermost regions, suggesting that ``downsizing'' is a rather
  universal phenomenon.

\item Metallicity gradients, when detected, are always negative, such
  that the galaxy centres are more metal-rich.  This result fits with
  a picture in which star formation continued in the central regions,
  with enriched material, after it had stopped in the outskirts.

\item For the brightest galaxies, gradients in age and metallicity are
  anti-correlated. This trend cannot be attributed to correlated errors. 
  It indicates that large differences in
  star formation history between the inner and outer parts of S0
  galaxies yield large differences in chemical enrichment.
\end{enumerate}

In combination with the previous results obtained in this study
(Papers\,I, II and III), a coherent picture is emerging in which S0
galaxies can be interpreted as spiral galaxies in which star formation
has ceased.  This conversion process seems to have occurred earliest in
the most massive galaxies as a further manifestation of downsizing.
What we have been able to add by studying the radial gradients in
these systems is the evidence that this cessation happened first in
the disks of these systems, with star formation shutting down from the
outside inward.  Such a sequence fits well with the requirement for a
process that enhances the central bulges of S0s relative to their
disks when compared to unquenched spiral systems.  

\section*{Acknowledgments}
This work was based on observations made with ESO telescopes at Paranal
Observatory under programme ID 070.A-0332. This publication makes use of data
products from the Two Micron All Sky Survey, which is a joint project of the
University of Massachusetts and the Infrared Processing and Analysis
Center/California Institute of Technology, funded by the National
Aeronautics and Space Administration and the National Science
Foundation.  This work has been supported by the Programa Nacional de Astronom\'{i}a y
Astrof\'{i}sica of the Spanish Ministry of Science and Innovation
under grants AYA2007-67752-C03-03 and AYA2006-15698-C02-02.

\appendix

\section{Line index gradients}

In Figs.\,\ref{fig:IndGradN1316} -- \ref{fig:IndGradE002}, radial measurements 
(along the semi-major axes) for 10 line indices are presented for the galaxy sample. 
The radii are scaled to the total galaxy half-light radius ($\rm R_{HALF}$) 
as calculated from $\rm K_{\rm s}$-band photometry. 
The radial profiles were folded around the galaxy centre 
(open and filled symbols; see each figure for details). 
Linear fits are shown for the radial index profiles (indices in magnitudes) with those points lying in the central region affected by seeing excluded from the fit. The vertical lines show different
radial scales for each galaxy when available. The dashed lines
(in green) mark the bulge effective radius ($\rm 1\,R_e$). The
vertical continuous lines (in red) mark the disk scale-length ($\rm
1\,R_d$). The dot-dashed lines (in blue) correspond to the radius
($\rm R_{LIM}$) from which the light profile is dominated by the disk
component as calculated in Paper\,I. When $\rm R_{LIM}$ is
accompanied by horizontal arrows, it indicates that the line position
is indicative only as its real location is too close to the galaxy
centre to be shown. The resulting values of the index
gradients and their uncertainties are presented in
Table\,\ref{tab:GradInd}.

\begin{figure*}
\begin{center}
\includegraphics[scale=.88]{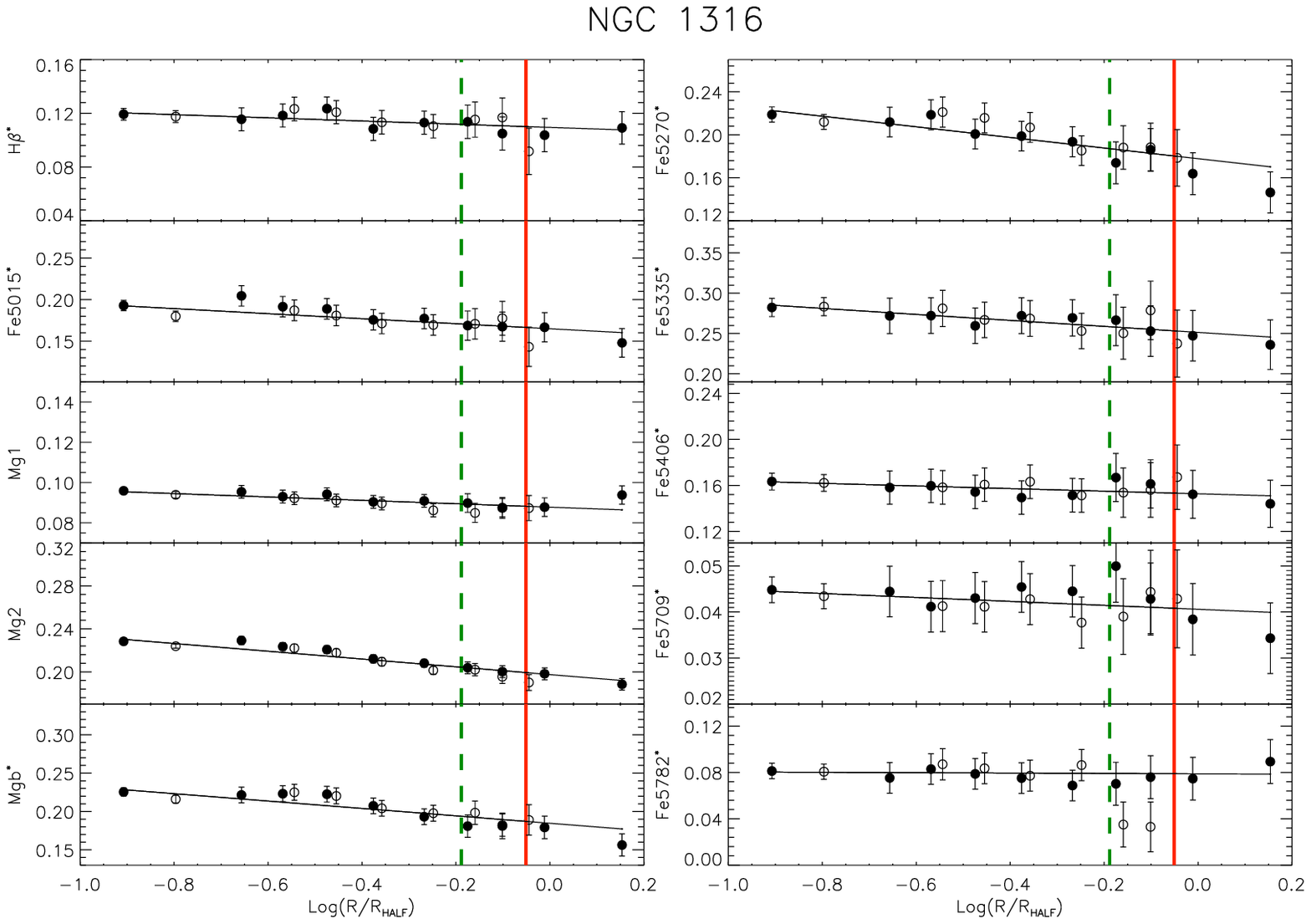}
\includegraphics[scale=.88]{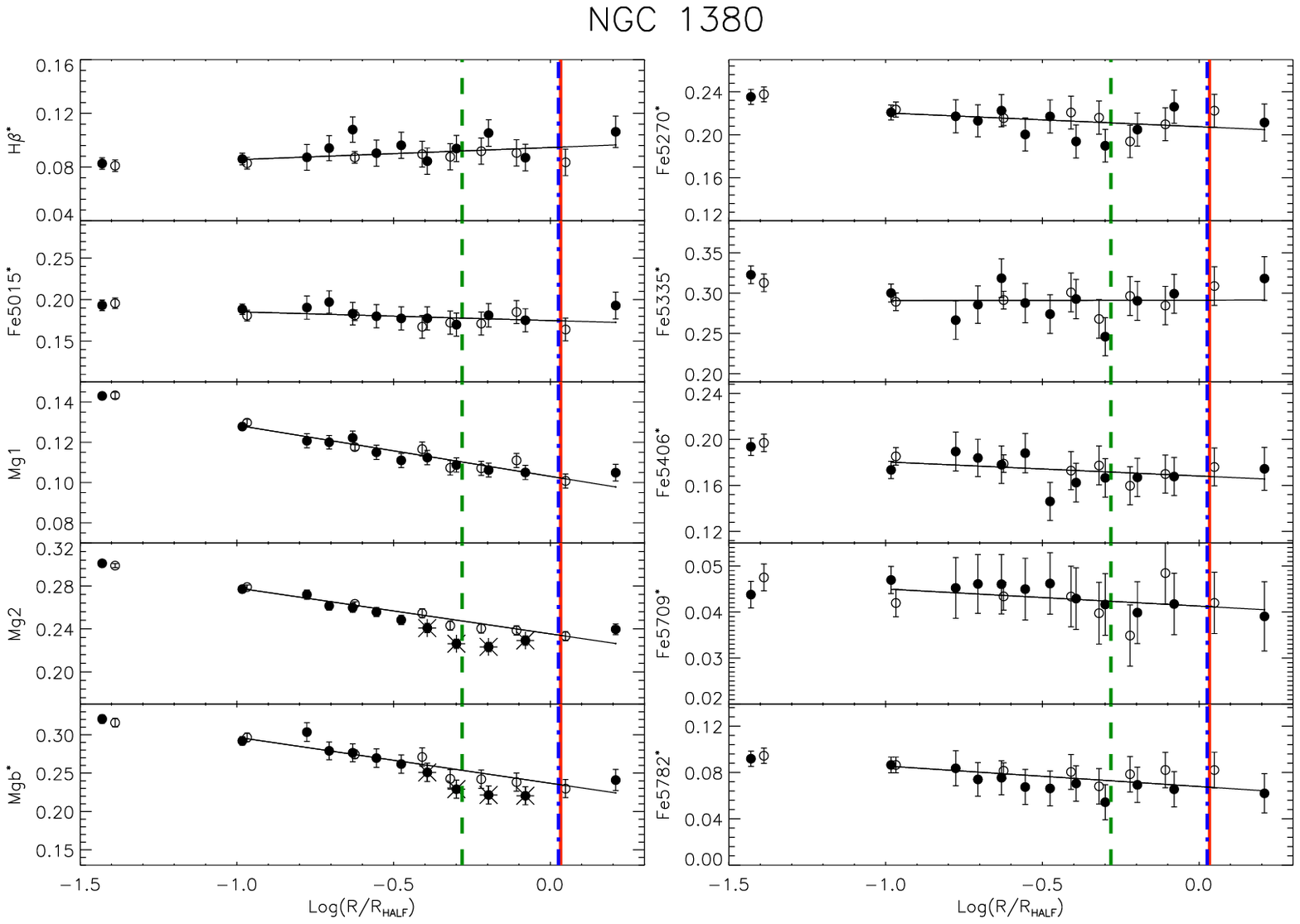}
\end{center}
\caption{\label{fig:IndGradN1316} Ten index gradients along the semi-major axis of NGC\,1316 and NGC\,1380. Filled (open) symbols correspond to: SW (NE) (for NGC\,1316) and S (N) (for NGC\,1380). Central bin indices are not shown. Linear fits (error-weighted gradients) are shown for the radial index profiles where points within seeing-size from the nucleus where not used. Points with ``*'' symbols in $\rm Mg_2$ and Mg$b$ indices where not used in the fit as they lie in a bad-pixel area of the detector. See the Appendix text for a description of the vertical lines.}
\end{figure*}

\begin{figure*}
\begin{center}
\includegraphics[scale=.9]{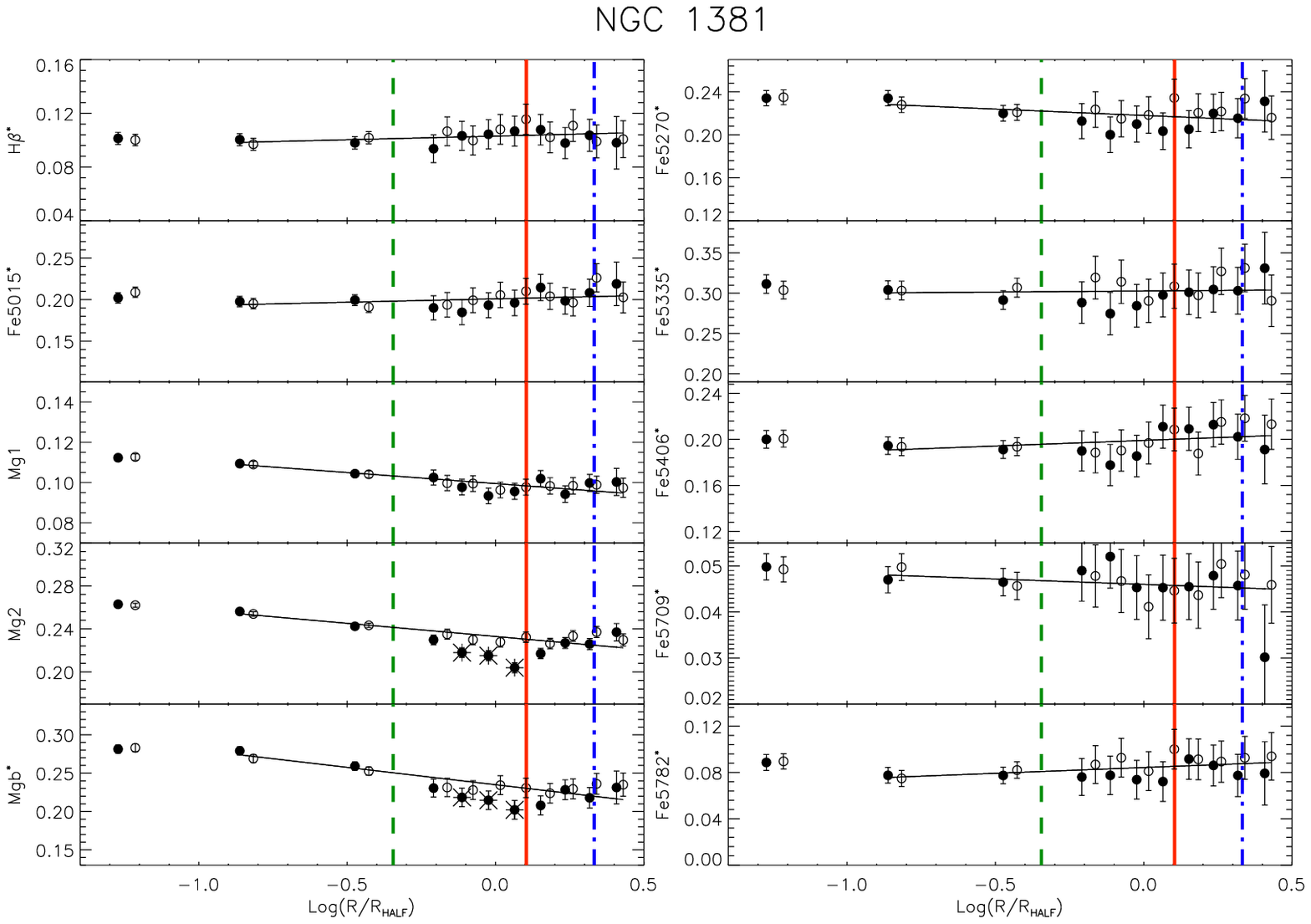}
\includegraphics[scale=.9]{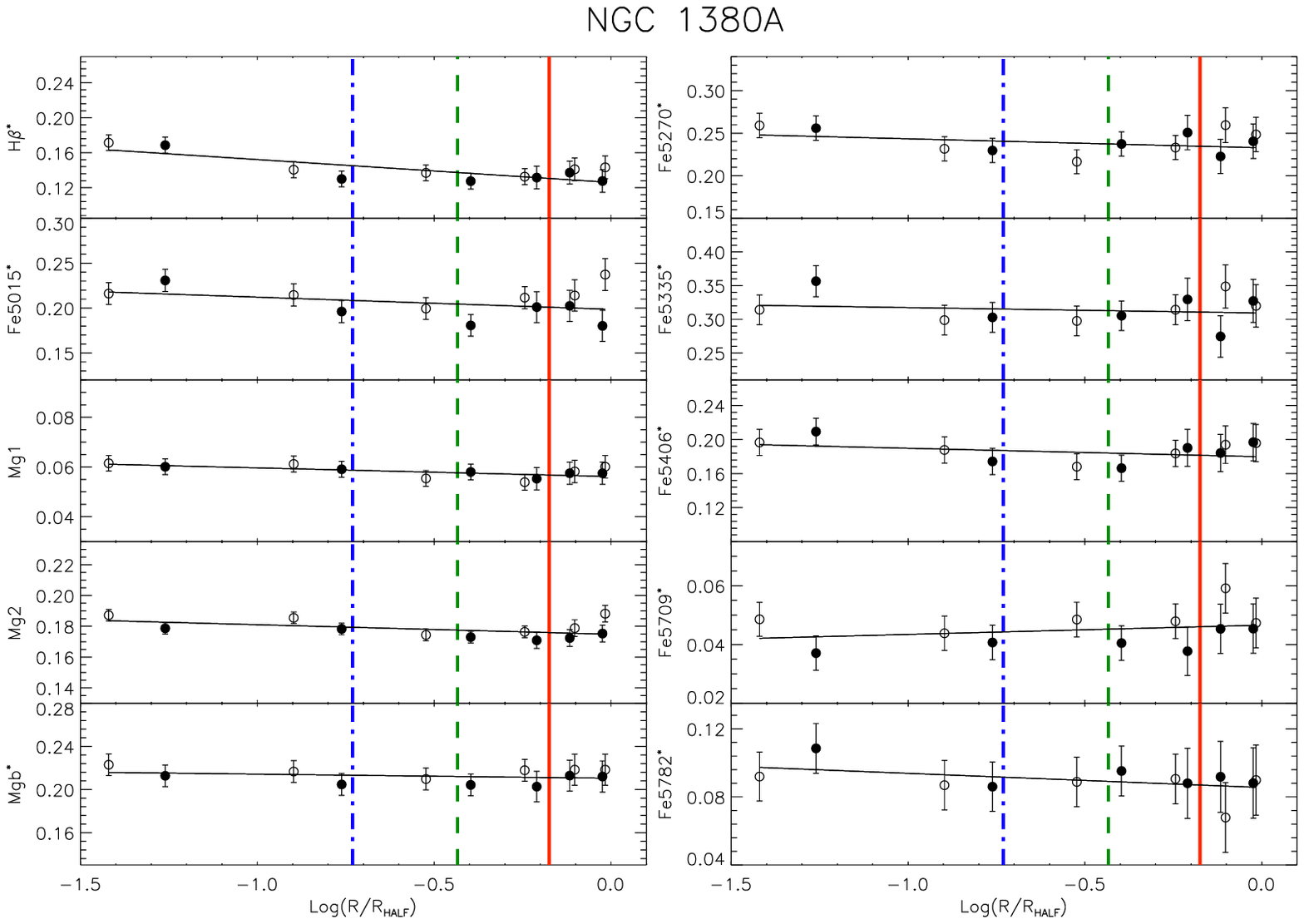}
\end{center}
\caption{\label{fig:IndGradN1381} Same as Fig.~\ref{fig:IndGradN1316} for NGC\,1381 and NGC\,1380A. Filled (open) symbols correspond to; SE (NW) (for NGC\,1381) and S (N) (for NGC\,1380A). Points with ``*'' symbols in Mg$_2$ and Mg$b$ indices where not used in the fit as they lie in a bad-pixel area of the detector.}
\end{figure*}

\begin{figure*}
\begin{center}
\includegraphics[scale=.9]{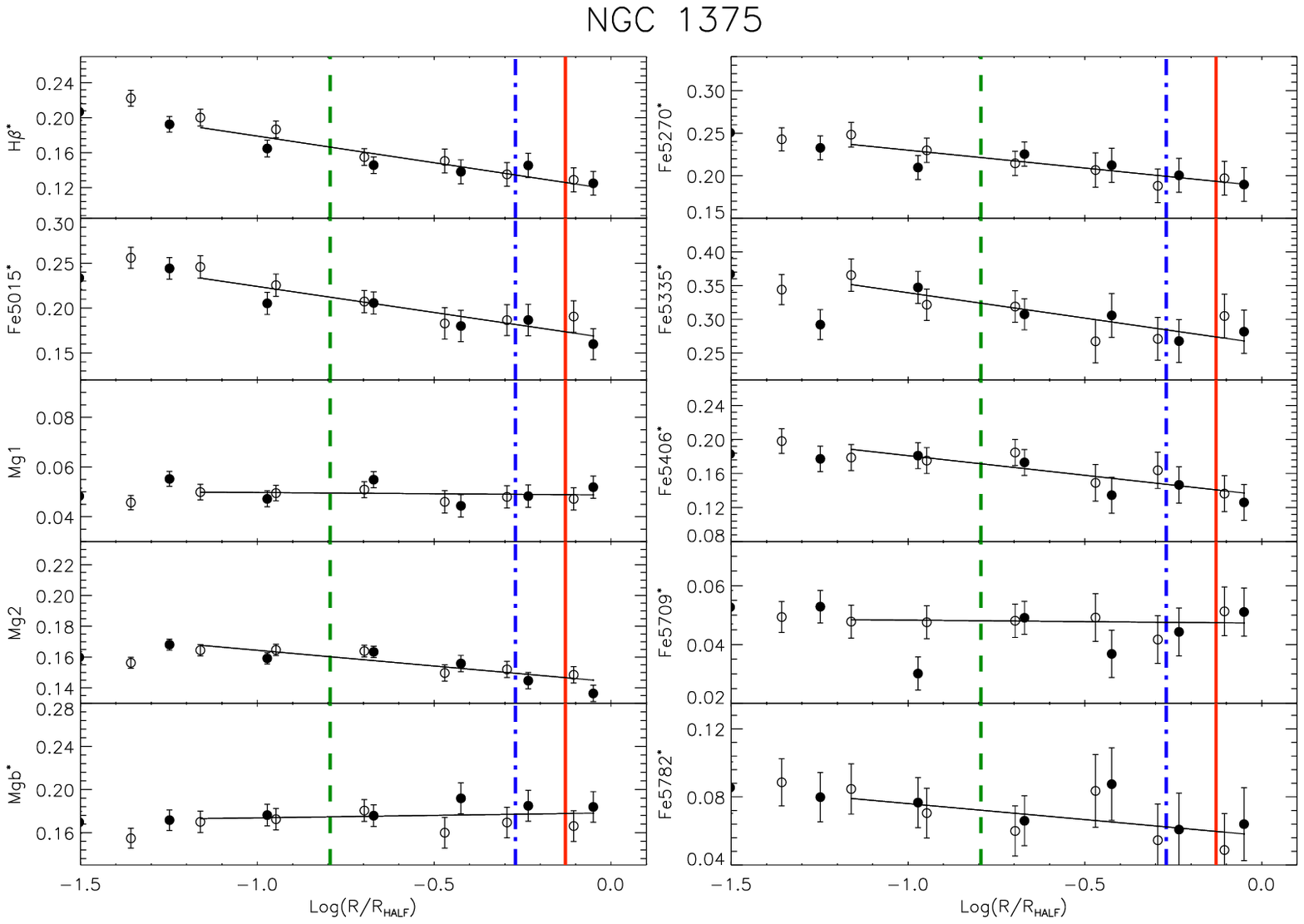}
\includegraphics[scale=.9]{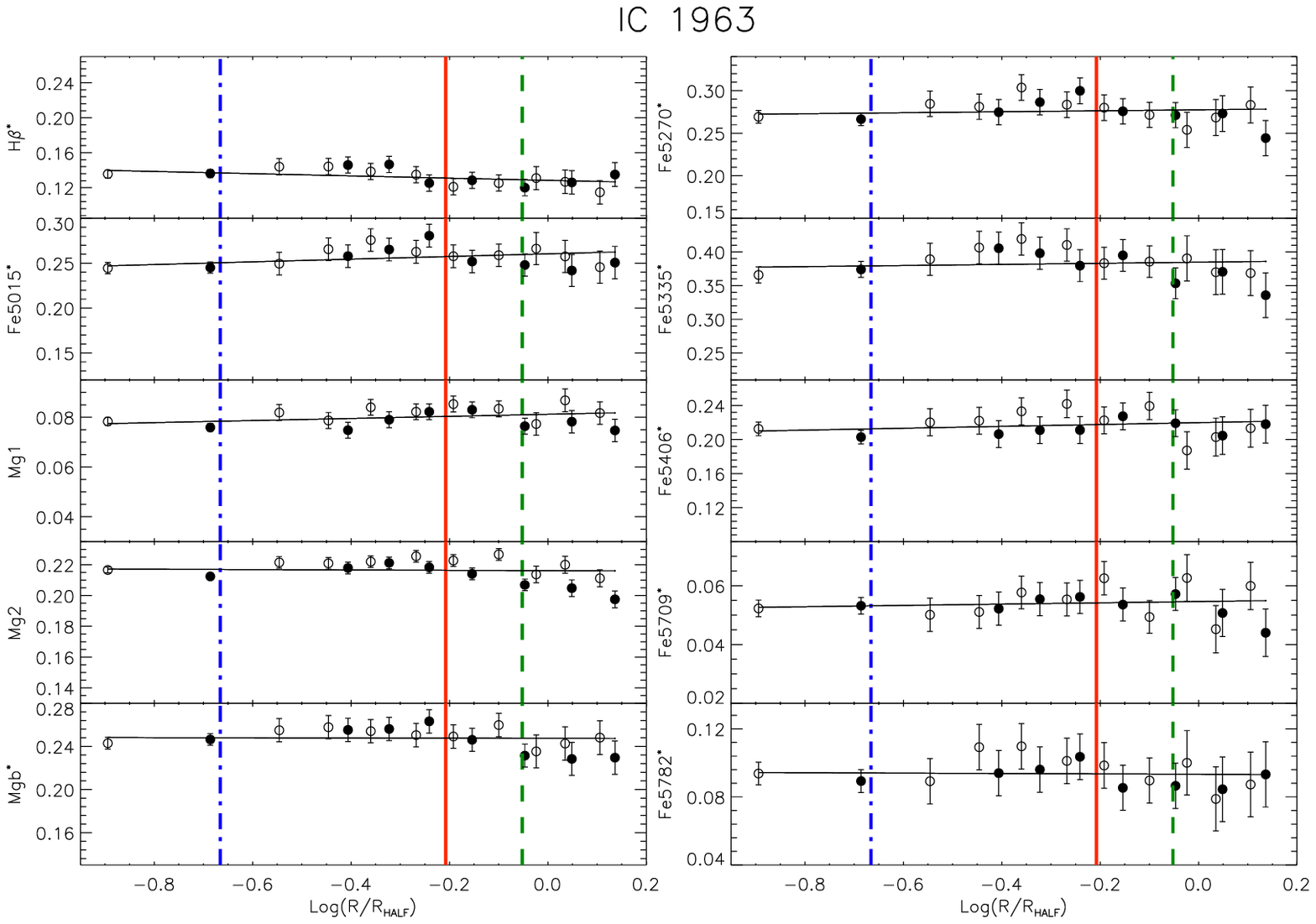}
\end{center}
\caption{\label{fig:IndGradN1375} Same as Fig.~\ref{fig:IndGradN1316} for NGC\,1375 and IC\,1963. Filled (open) symbols correspond to: the W (E) profile projected in the sky (for both galaxies).}
\end{figure*}

\begin{figure*}
\begin{center}
\includegraphics[scale=.9]{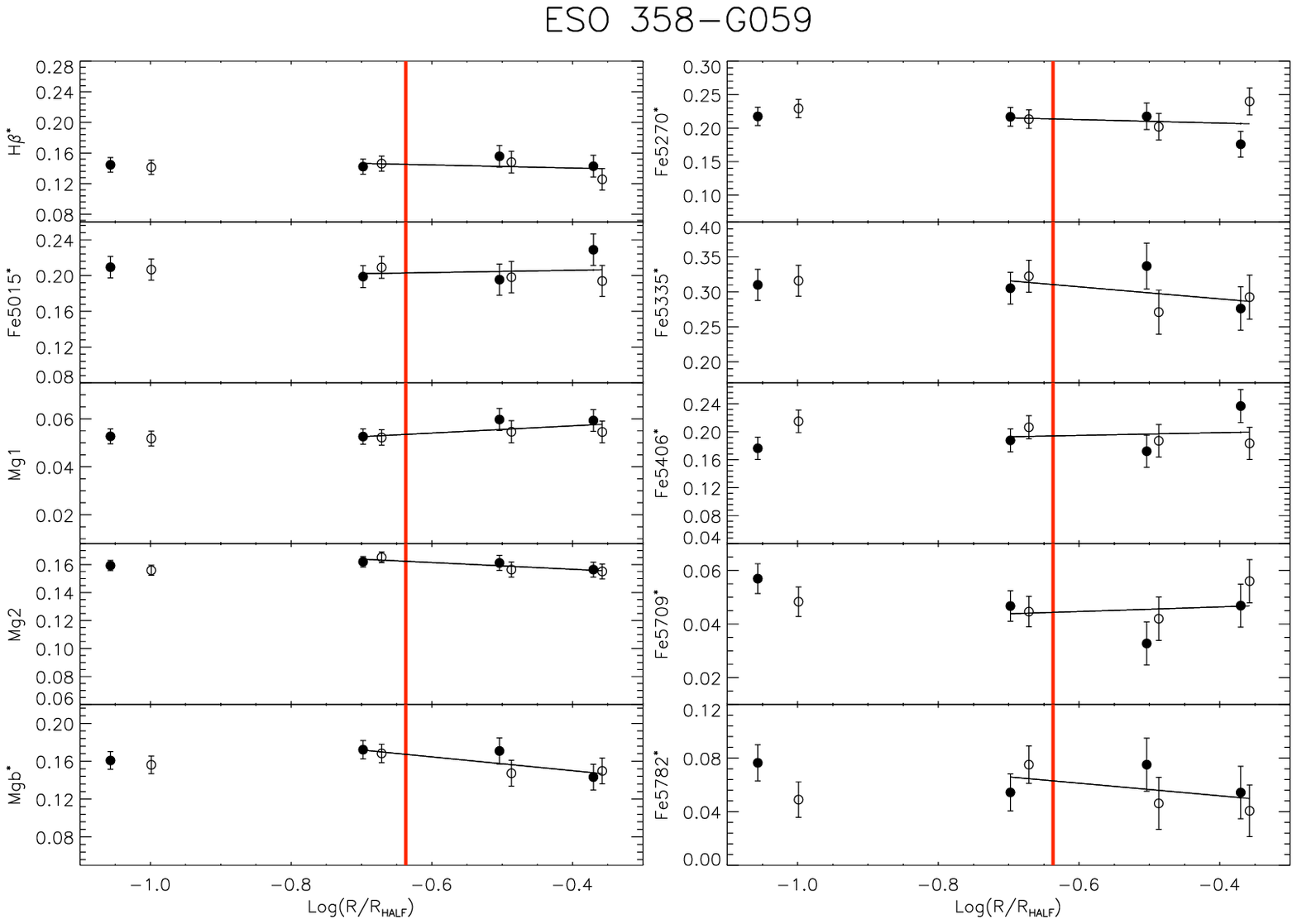}
\includegraphics[scale=.9]{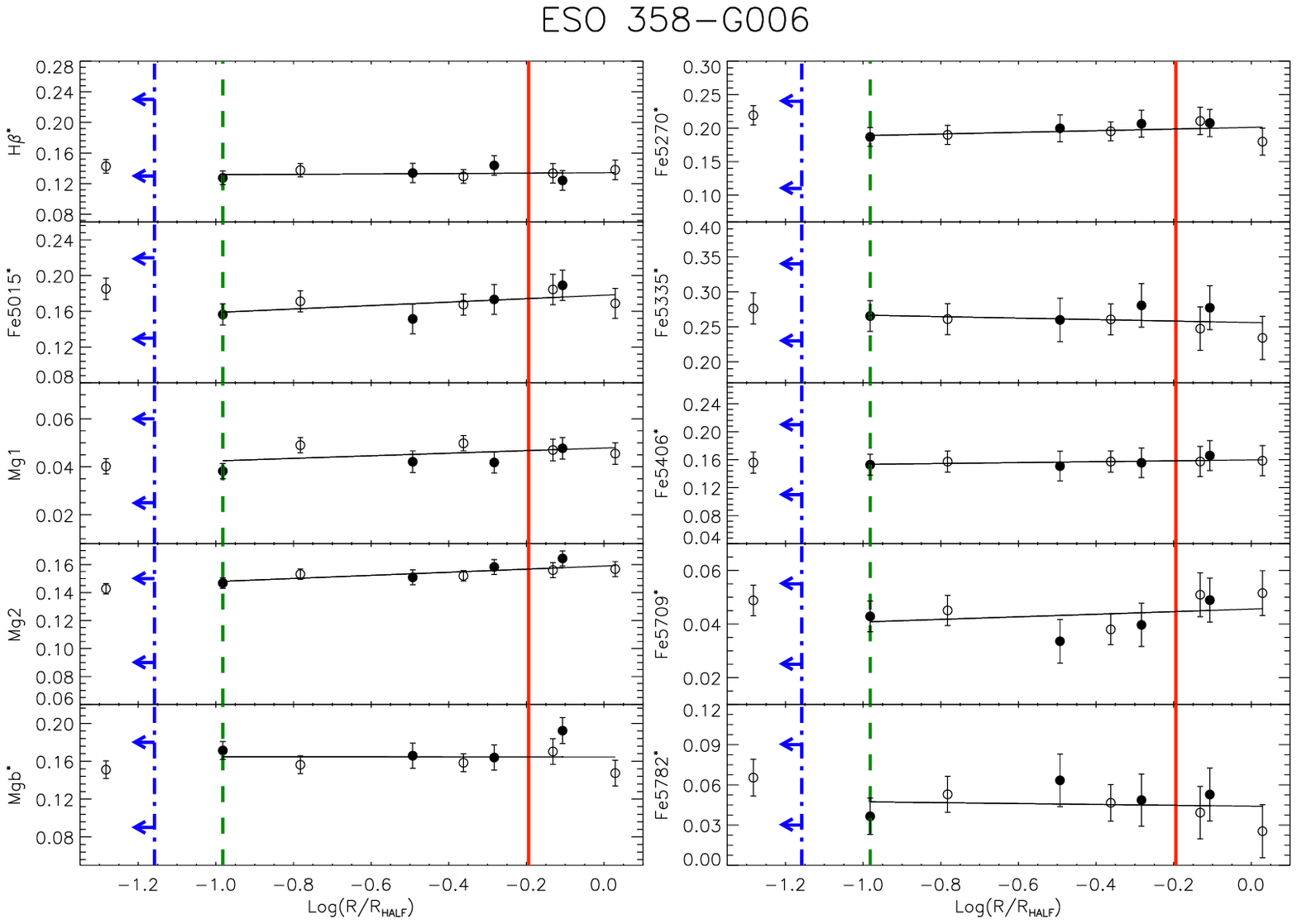}
\end{center}
\caption{\label{fig:IndGradE059} Same as Fig.~\ref{fig:IndGradN1316} for ESO\,358-G059 and ESO\,358-G006. Filled (open) symbols correspond to the SE (NW) (for ESO\,358-G059) and SW (NE) (for ESO\,358-G006).}
\end{figure*}

\begin{figure*}
\begin{center}
\includegraphics[scale=.9]{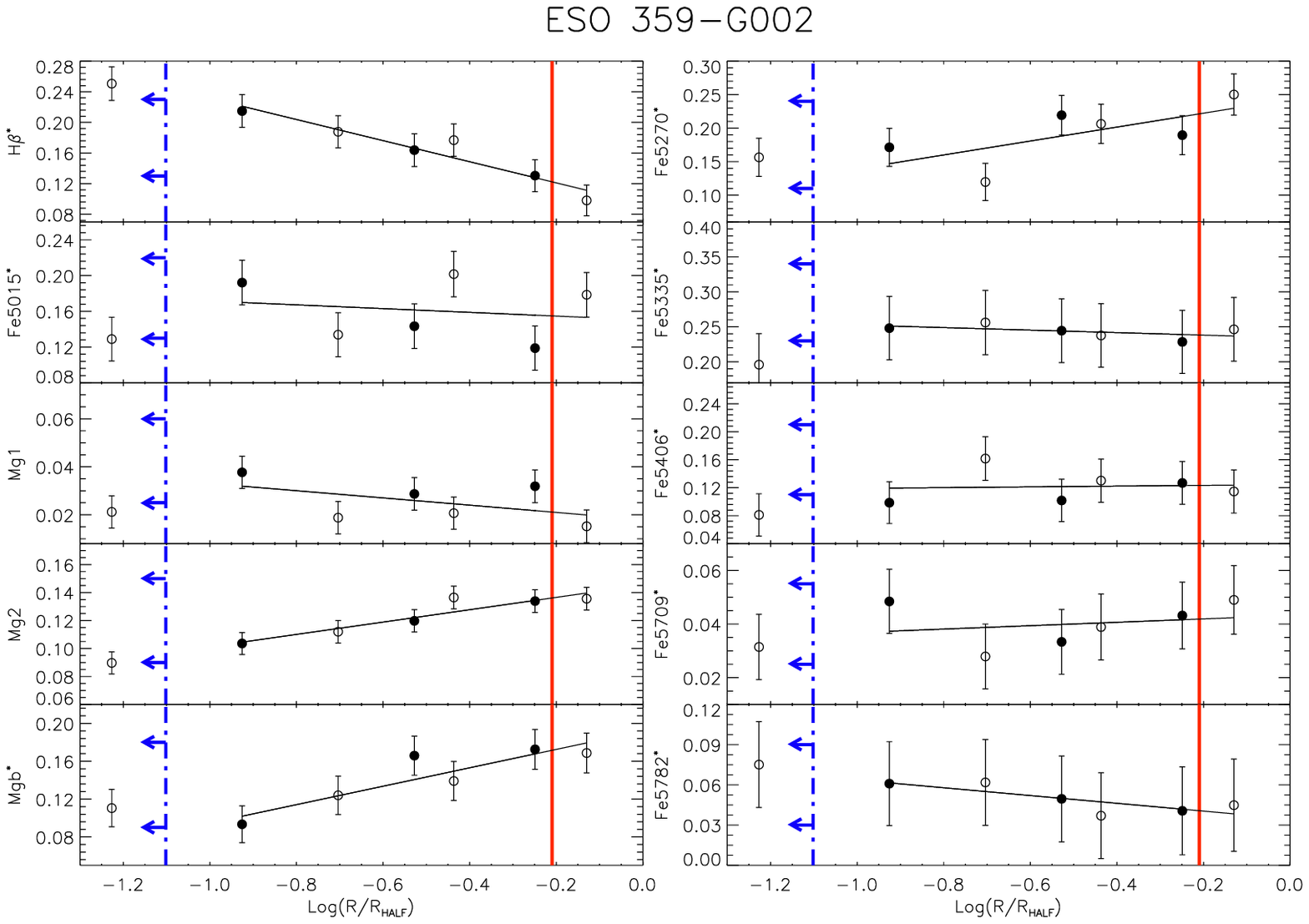}
\end{center}
\caption{\label{fig:IndGradE002} Same as Fig.~\ref{fig:IndGradN1316} for ESO\,359-G002. Filled (open) symbols correspond to the SW (NE) profile projected in the sky.}
\end{figure*}

\begin{landscape}
\begin{table}
\begin{center}
\caption{\label{tab:GradInd} Line Index Gradients of S0 galaxies in Fornax Cluster.}
\begin{tabular}{@{}lc@{\,}c@{\,}c@{\,}c@{\,}c@{\,}c@{\,}c@{\,\,\,}c@{\,\,\,}c@{\,\,\,}c@{\,}c@{\,}c@{}}
\hline
\hline
 Name & Grad(H$\beta^*$) & Grad(Fe5015$^*$) & Grad($\rm Mg_1$) &     Grad($\rm Mg_2$) & Grad(Mg$b^*$) & Grad(Fe5270$^*$) & Grad(Fe5335$^*$) & Grad(Fe5406$^*$) & Grad(Fe5709$^*$) &  Grad(Fe5782$^*$) \\

    (1)   &      (2) &     (3)  &   (4)  &    (5)    &     (6)    &   (7)   &       (8) & (9) & (10) & (11)   \\
\hline
\hline
\\[0.2em]
   NGC\,1316   &    $-$0.012   (0.007)       &   {\bf $ -$0.030    (0.010) }     &      {\bf $-$0.008     (0.002) }     &    {\bf $-$0.036   (0.003) }      &    {\bf $-$0.048   (0.008) }     &     {\bf $-$0.049   (0.011)}       &   $-$0.037   (0.016)       &   $-$0.011   (0.011)    &$-$0.004   (0.004)  &  $-$0.002   (0.010)   \\\\
   NGC\,1380   &   0.009     (0.005)       &    $-$0.011    (0.008)    &      {\bf $-$0.026     (0.002)}      &    {\bf $-$0.043   (0.003)}      &    {\bf $-$0.060   (0.007)}     &     $-$0.013   (0.008)      &      0.001   (0.013)     &     $-$0.012   (0.009)   & $-$0.004   (0.003)  &  $-$0.018   (0.009)       \\\\
   NGC\,1381      &   0.005     (0.005)       &      0.008    (0.005)   &       $-$0.016     (0.006)      &     {\bf $-$0.025   (0.002)}     &      {\bf $-$0.045   (0.005)}    &      $-$0.012   (0.007)     &       0.003   (0.011)    &        0.010   (0.008)   & $-$0.002   (0.003)  &    0.010   (0.007)       \\\hline
   NGC\,1380A    &     {\bf $-$0.026   (0.007)}       &   $-$0.014    (0.009)     &     $-$0.004     (0.002)      &    $-$0.006   (0.003)      &    $-$0.004   (0.007)      &    $-$0.011   (0.010)       &   $-$0.008   (0.015)       &   $-$0.010   (0.011)   &   0.003   (0.004)  &  $-$0.008   (0.010)       \\\\
   NGC\,1375   &     {\bf $-$0.061   (0.010)}       &    {\bf $-$0.058    (0.012)}     &     $-$0.001     (0.003)      &     {\bf $-$0.020   (0.004) }     &      0.005   (0.010)     &      {\bf $-$0.042   (0.014)}      &     {\bf $-$0.075   (0.022)}      &    $-$0.046   (0.016)   &   0.005   (0.006)  &  $-$0.019   (0.015)       \\\\
  IC\,1963     &    $-$0.013   (0.006)       &     0.015    (0.009)    &        0.000     (0.005)      &    $-$0.002   (0.003)     &     $-$0.001  (0.007)    &        0.006   (0.010)    &        0.008   (0.016)   &         0.011   (0.010)   &   0.002   (0.004)   & $-$0.001   (0.009)         \\\hline
 {\scriptsize ESO\,358-G059}    &    $-$0.020   (0.035)       &     0.013    (0.047)    &        0.018     (0.014)      &    $-$0.024   (0.013)     &     $-$0.073   (0.037)    &      $-$0.026   (0.051)     &     $-$0.086   (0.081)     &       0.020   (0.058)   &   0.009   (0.021)   & $-$0.047   (0.050)       \\\\
 {\scriptsize ESO\,358-G006}     &   0.003     (0.011)       &      0.014    (0.019)   &          {\bf  0.047     (0.000)}      &      0.018   (0.070)   &         0.000         (0.013)  &          0.012   (0.018)   &       $-$0.011   (0.028)   &         0.006   (0.019)   &   0.005   (0.007)   & $-$0.003   (0.017)           \\\\
 {\scriptsize ESO\,359-G002}     &     {\bf $-$0.138   (0.031)}       &   $-$0.021    (0.039)     &     $-$0.015     (0.010)      &       {\bf  0.044   (0.012)}      &       {\bf  0.097   (0.032)}    &         {\bf  0.104   (0.046)}    &      $-$0.018   (0.073)    &        0.005   (0.048)   &   0.006   (0.019)   & $-$0.029   (0.049)       \\[0.5em]
\hline
\hline
\end{tabular}\\
\end{center}
\footnotesize{Notes: From (2) to (11), $1\,\sigma$ RMS errors
  between $\rm ^{"}(\,)^{"}$; Col (1), galaxy name; Col (2), H$\beta$ index; Col (3), Fe5015 index;
  Col (4), $\rm Mg_1$ index in magnitudes; Col (5), $\rm Mg_2$ index in magnitudes; Col (6),
  Mg$b$ index; Col (7), Fe5270 index; Col (8), Fe5335 index; Col (9), Fe5406
  index; Col (10), Fe5709 index; Col (11), Fe5782 index. Gradients calculated with line indices 
in magnitudes (Paper~III) with respect to $\rm \log(R/R_{HALF})$ (radii normalized by the half-light radius of each galaxy).
Significant gradients ($\ge 3\,\sigma$ detections) are highlighted in {\bf bold}.}
\end{table}
\end{landscape}

\section{SSP parameters}

Figs.\,\ref{fig:PGradN1316} -- \ref{fig:PGradE002} show SSP ages and
[M/H] derived using Vazdekis et al.\,(2010) models, and the
$\alpha$-element abundance tracer (Mg$b$/$\rm \langle Fe\rangle$
ratio).  The notation ``$\log{(\rm Age[yr])\ \ ([MgFe]'}$ --
H$\beta$)'' and ``${\rm [M/H]\ \ ([MgFe]'}$ -- H$\beta$)'' is used to
make explicit the fact that age and [M/H] have been derived using
[MgFe]' -- H$\beta$ diagrams such as the one in
Fig.\,\ref{fig:n1316_3MILES}, left panel.  The solid lines (in black)
show the best linear fits to the radial profiles.  The
slopes of these fits and corresponding errors are given in
Table~\ref{tab:GradAZA}.

The horizontal dashed lines in [M/H] plots represent solar
metallicity. The two horizontal dashed lines in the
[$\alpha$/Fe]-tracer plots delimit the log(Mg$b$/$\rm \langle
Fe\rangle$) range corresponding to [$\alpha$/Fe]=0 [for metallicities
$\rm -1.35 \le [Z/H] \le 0.35$ and ages between $\rm 3\,Gyr \le t \le
15\,Gyr$ according to Thomas, Maraston \& Bender (2003) models; for
NGC\,1380 and 1381 the range is limited to models of ages between $\rm
8\,Gyr \le t \le 15\,Gyr$]. Horizontal lines as in Figs.\,\ref{fig:IndGradN1316} -- \ref{fig:IndGradE002}.

\begin{figure*}
\begin{center}
\includegraphics[scale=.88]{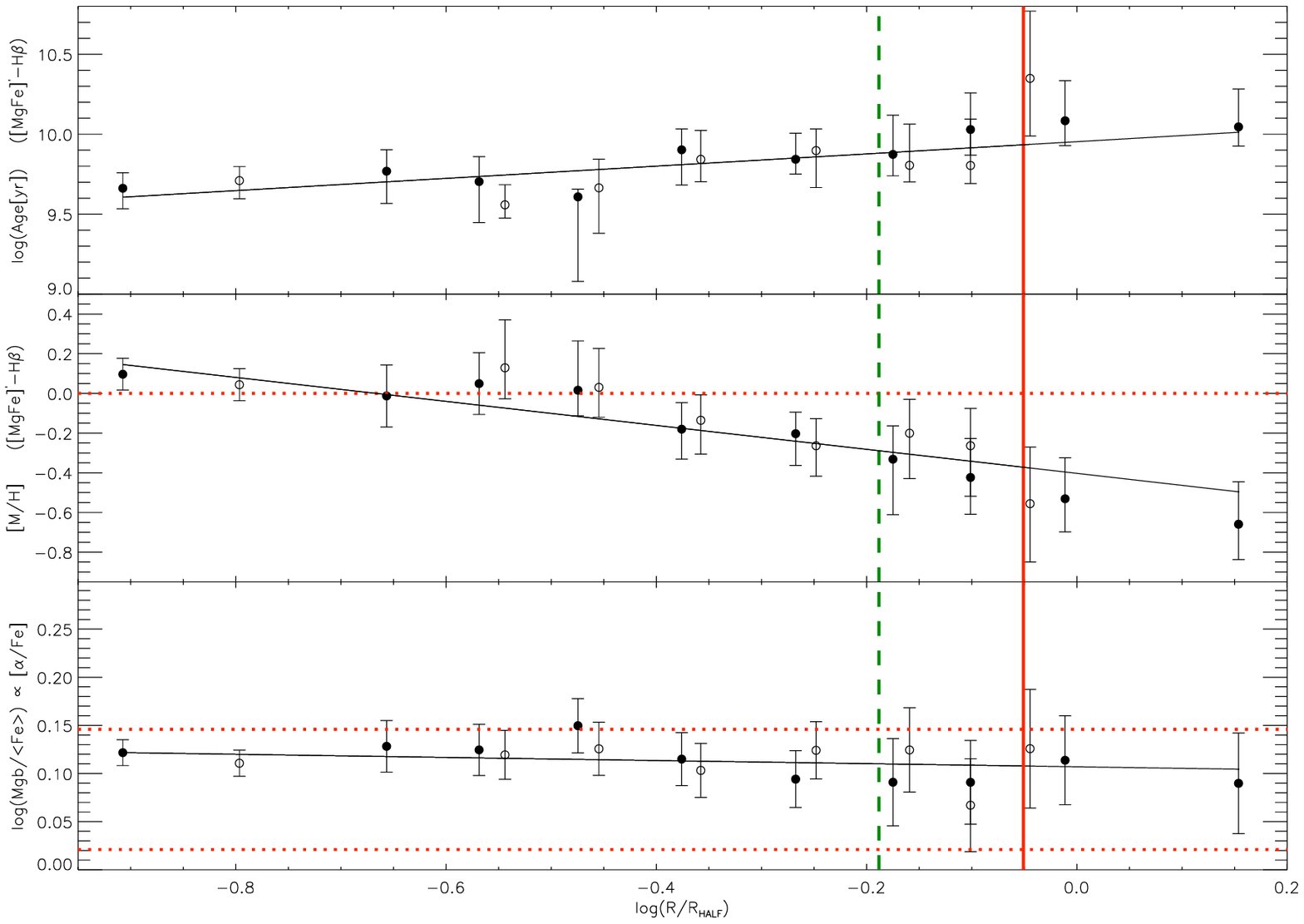}
\includegraphics[scale=.88]{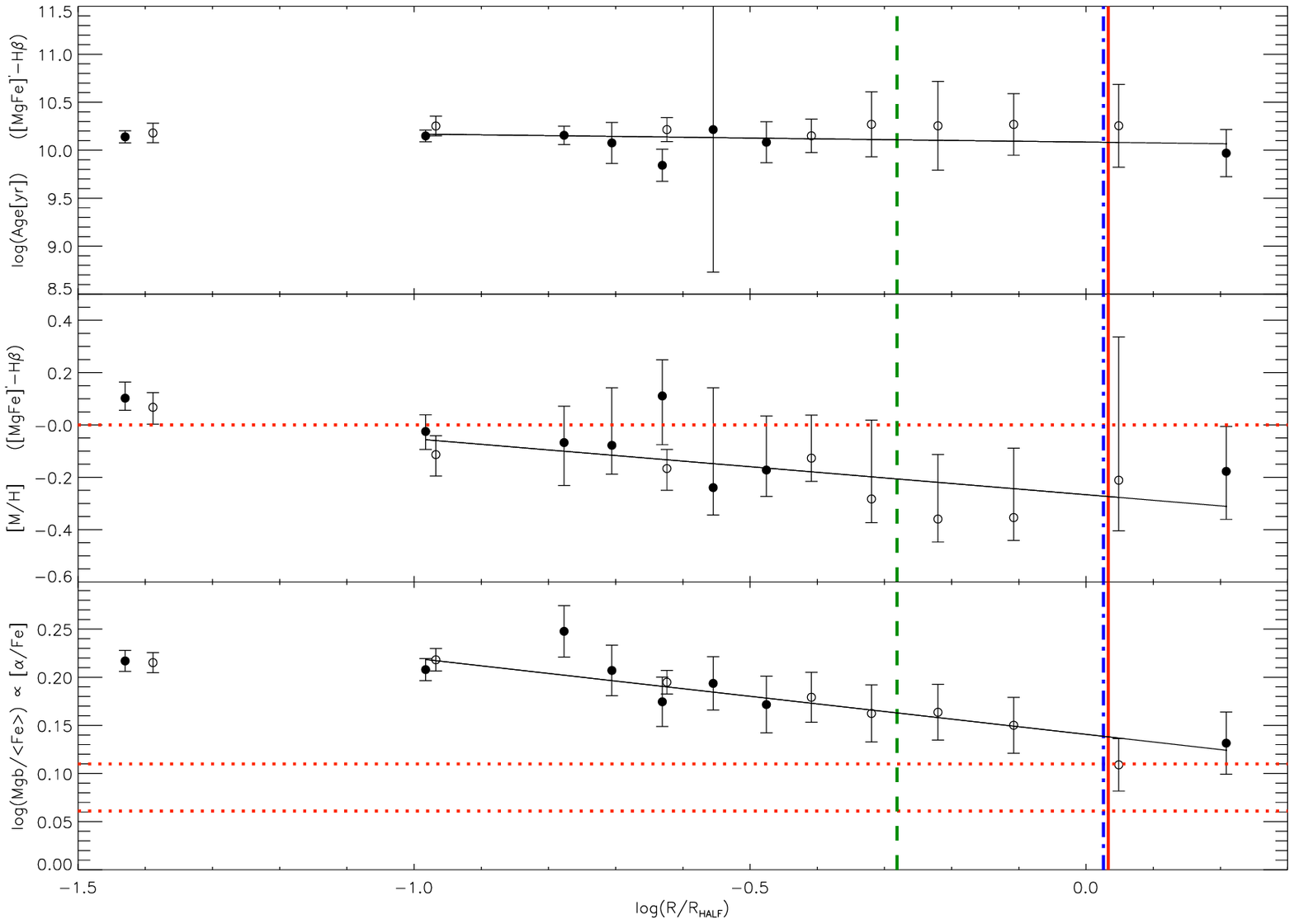}
\end{center}
\caption{\label{fig:PGradN1316} Gradients along the semi-major axis of NGC\,1316 (top) and NGC\,1380 (bottom), including predicted Ages, Metallicities and $\alpha$-element tracer values. Filled (open) symbols correspond to: SW (NE) (for NGC\,1316) and S (N) (for NGC\,1380). The description of the different lines can be found in the text. 
}
\end{figure*}

\begin{figure*}
\begin{center}
\includegraphics[scale=.88]{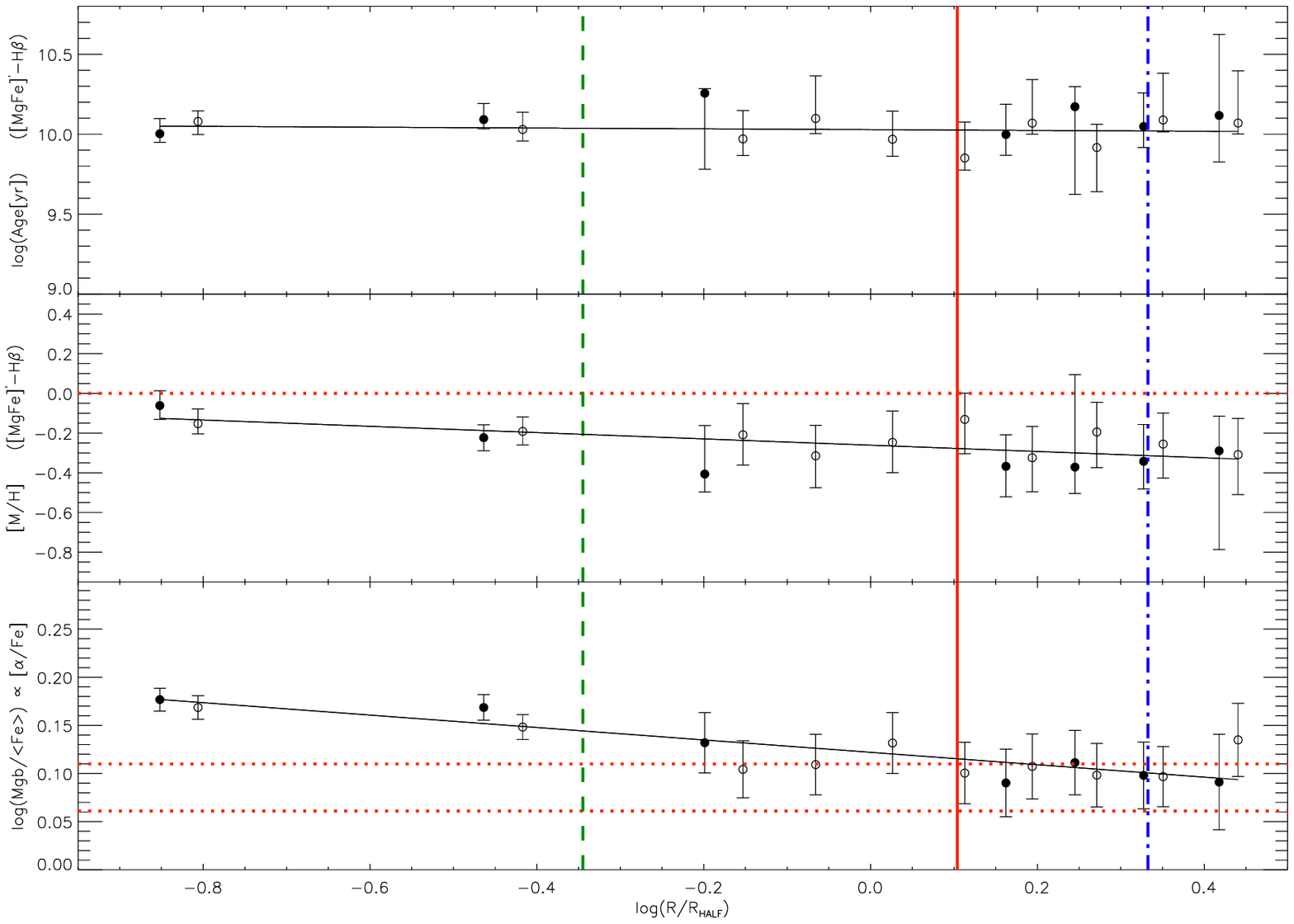}
\includegraphics[scale=.88]{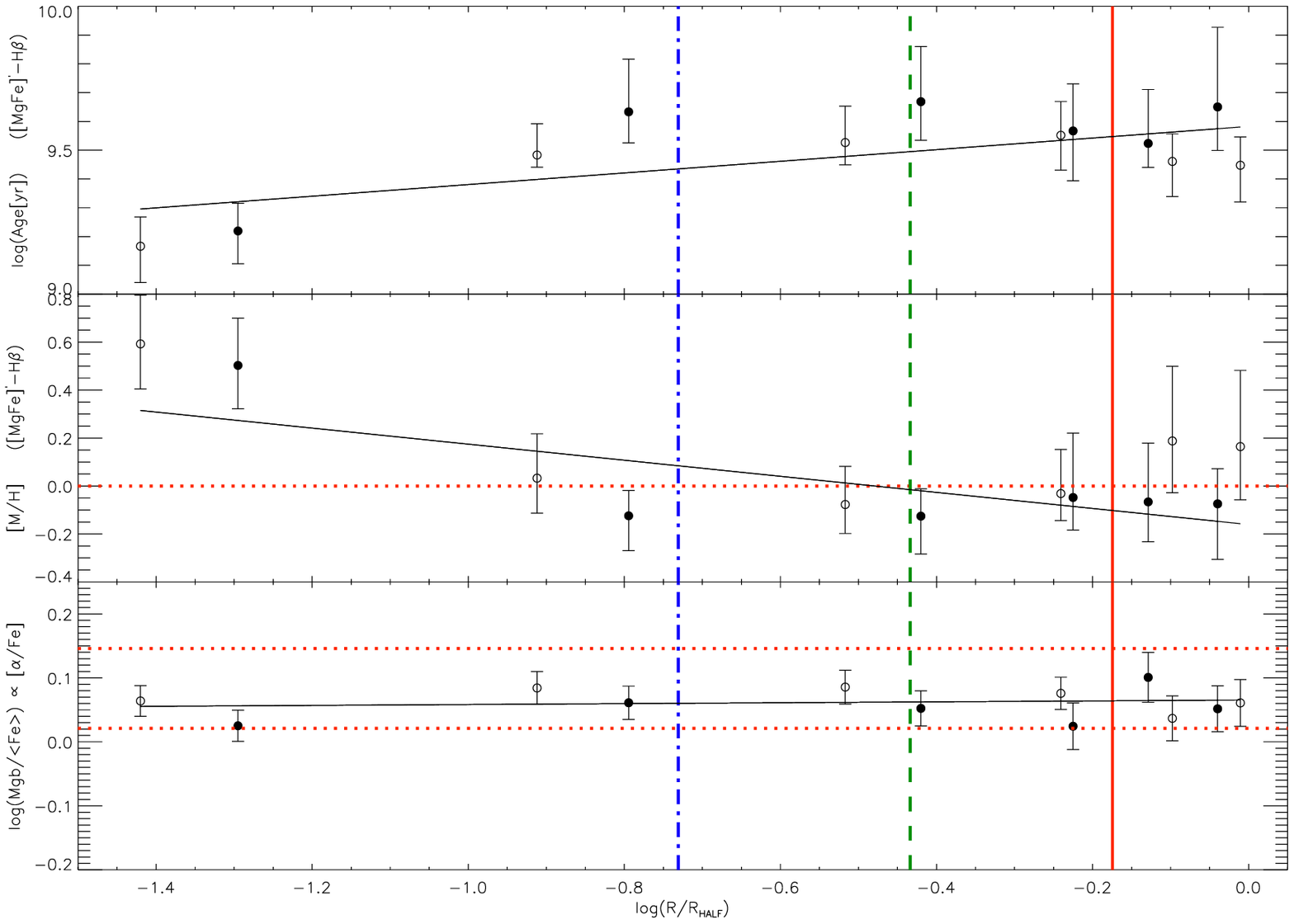}
\end{center}
\caption{\label{fig:PGradN1381} Same as Fig.~\ref{fig:PGradN1316} for NGC\,1381 (top) and NGC\,1380A (bottom). Filled (open) symbols correspond to; SE (NW) (for NGC\,1381) and S (N) (for NGC\,1380A).}
\end{figure*}

\begin{figure*}
\begin{center}
\includegraphics[scale=.88]{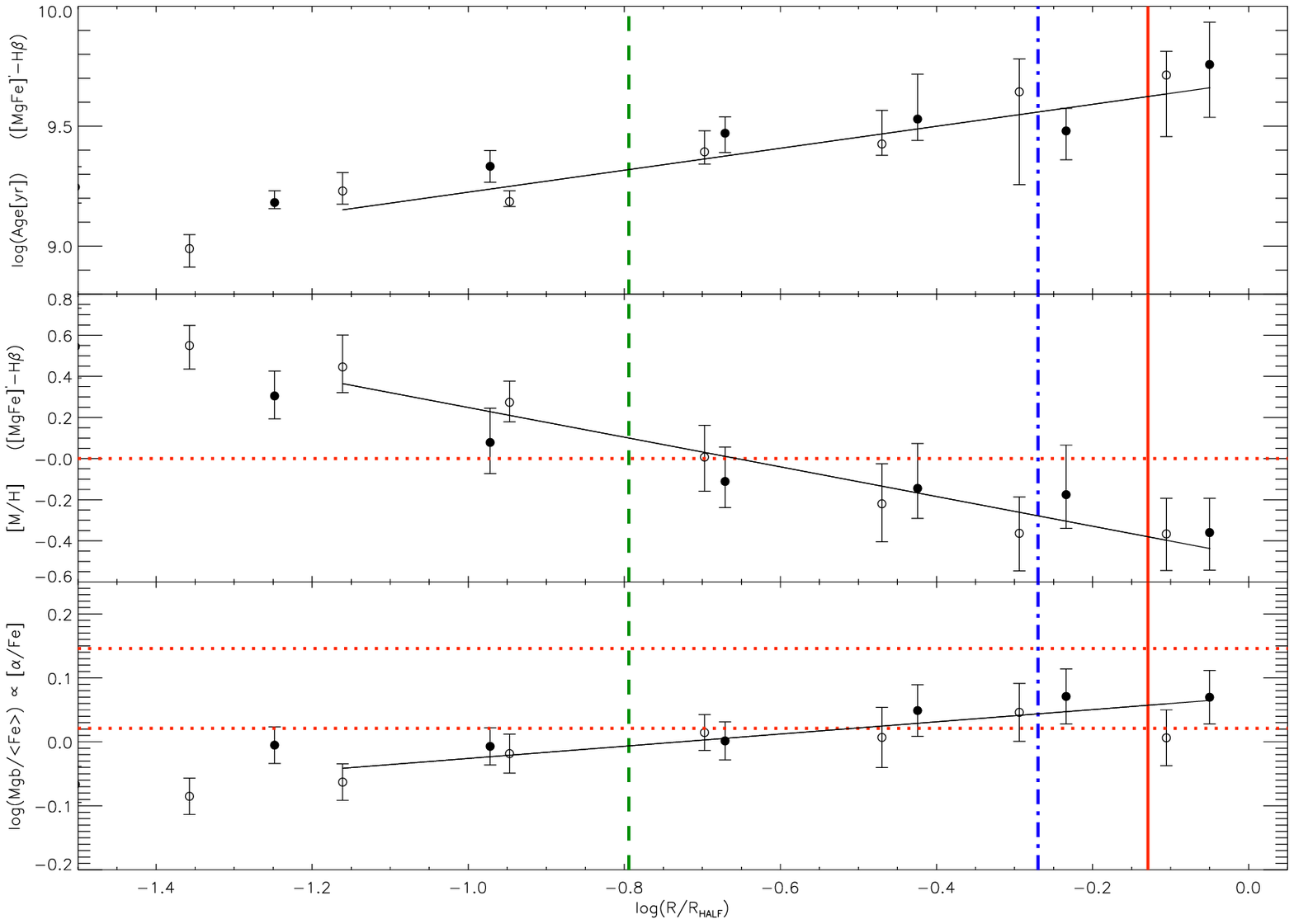}
\includegraphics[scale=.88]{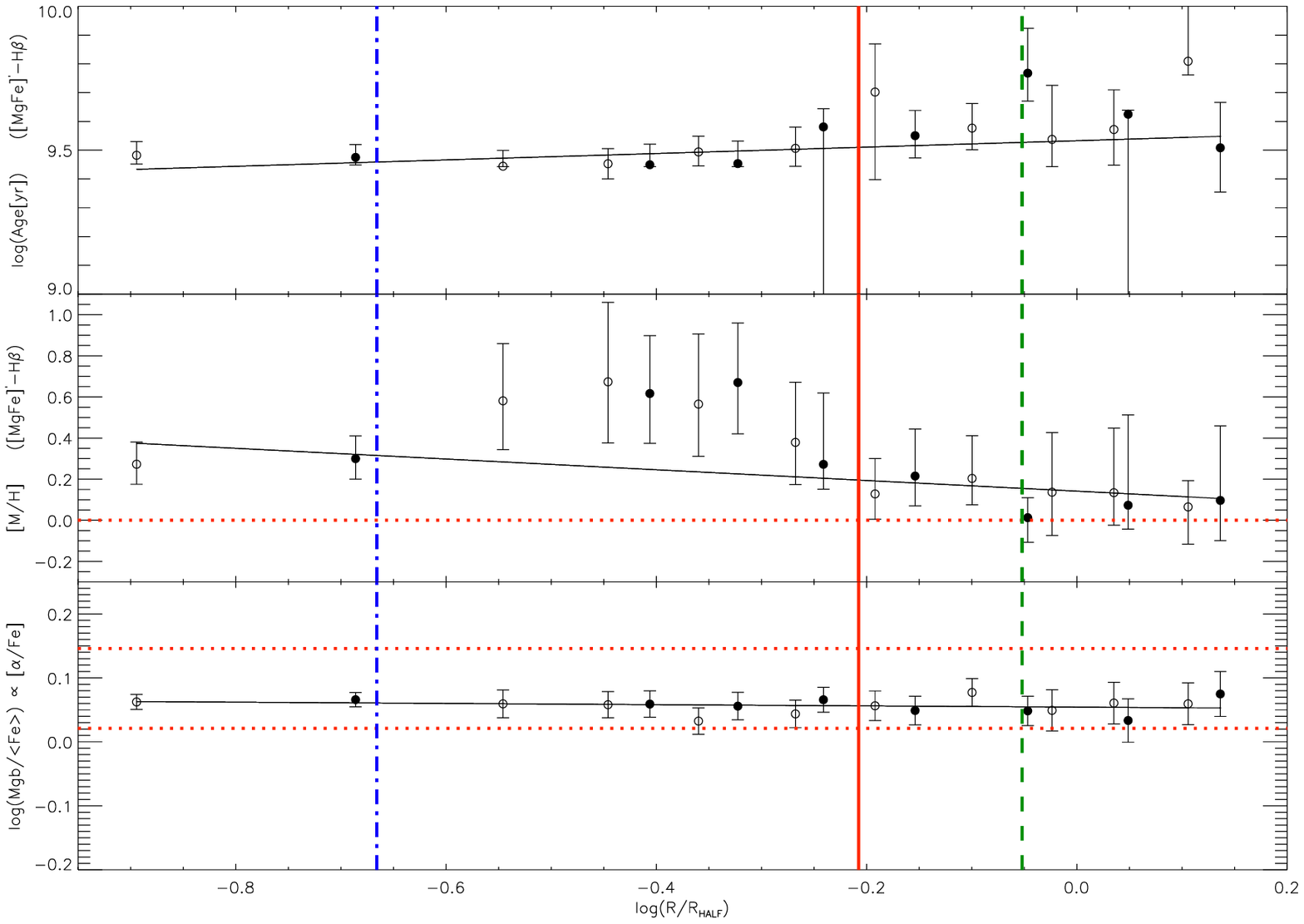}
\end{center}
\caption{\label{fig:PGradN1375} Same as Fig.~\ref{fig:PGradN1316} for NGC\,1375 (top) and  IC\,1963 (bottom). Filled (open) symbols correspond to: the W (E) profile projected in the sky (for both galaxies).}
\end{figure*}

\begin{figure*}
\begin{center}
\includegraphics[scale=.88]{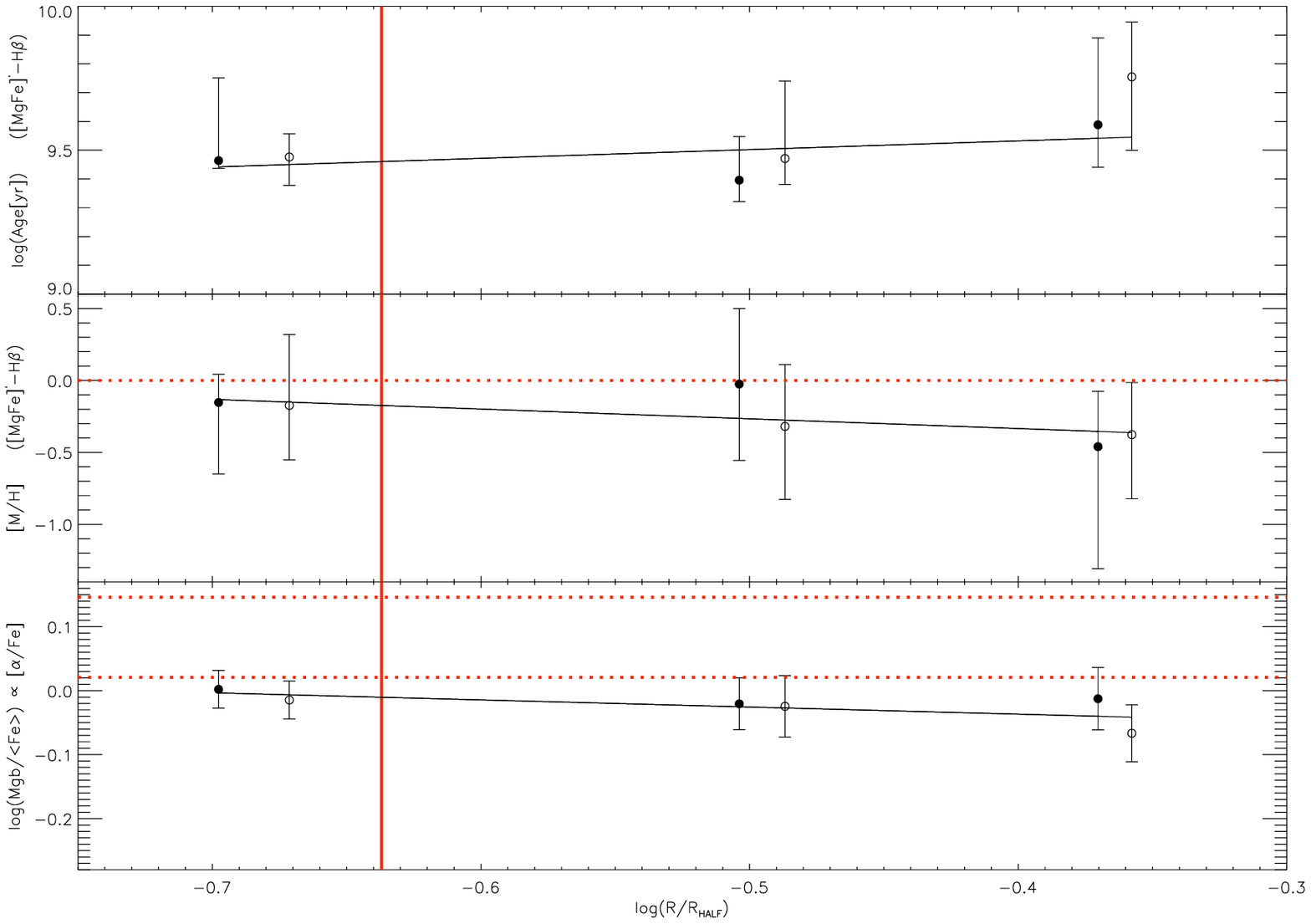}
\includegraphics[scale=.88]{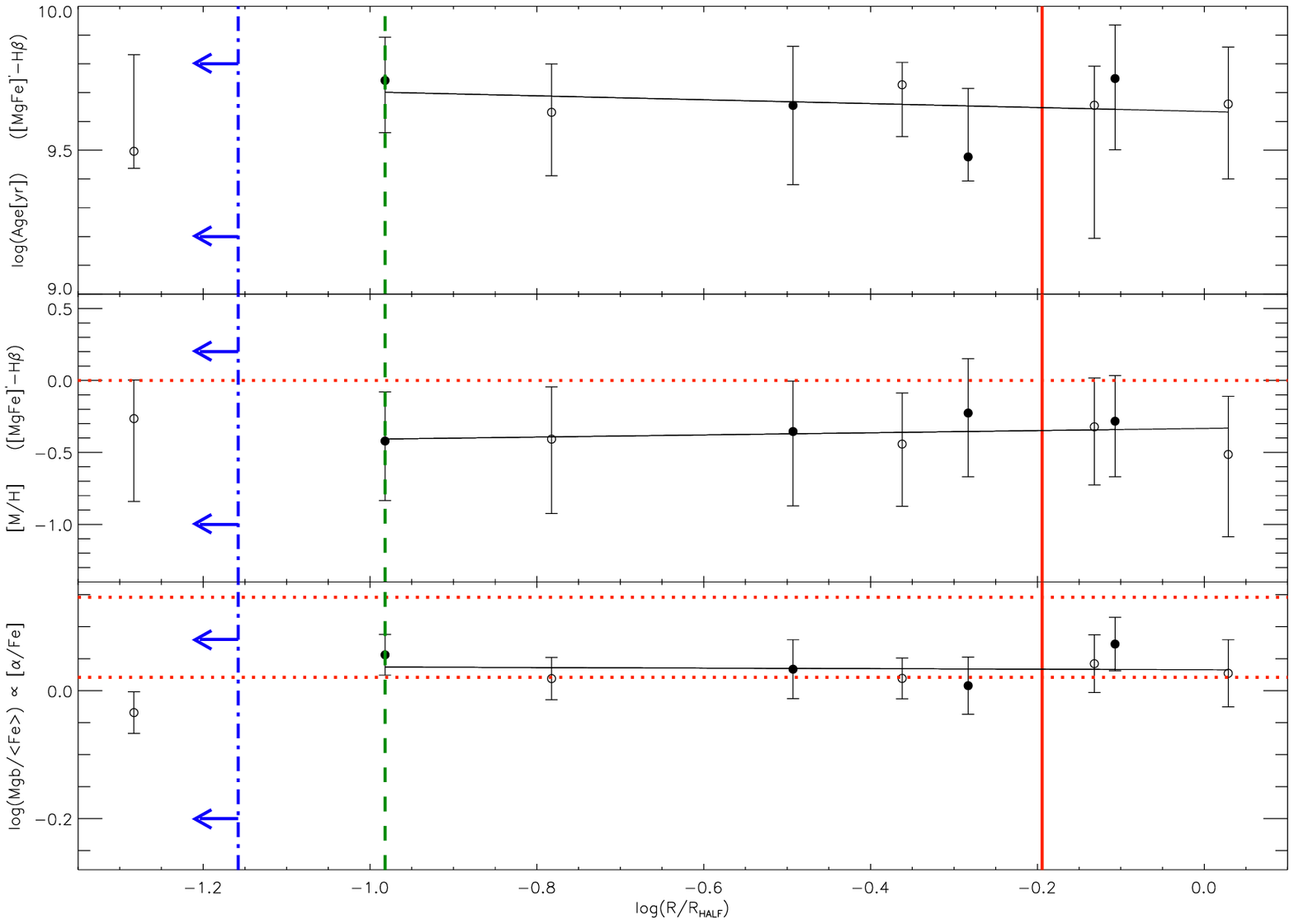}
\end{center}
\caption{\label{fig:PGradE059} Same as Fig.~\ref{fig:PGradN1316} for ESO\,358-G059 (top) and ESO\,358-G006 (bottom). Filled (open) symbols correspond to the SE (NW) (for ESO\,358-G059) and SW (NE) (for ESO\,358-G006).}
\end{figure*}

\begin{figure*}
\begin{center}
\includegraphics[scale=.88]{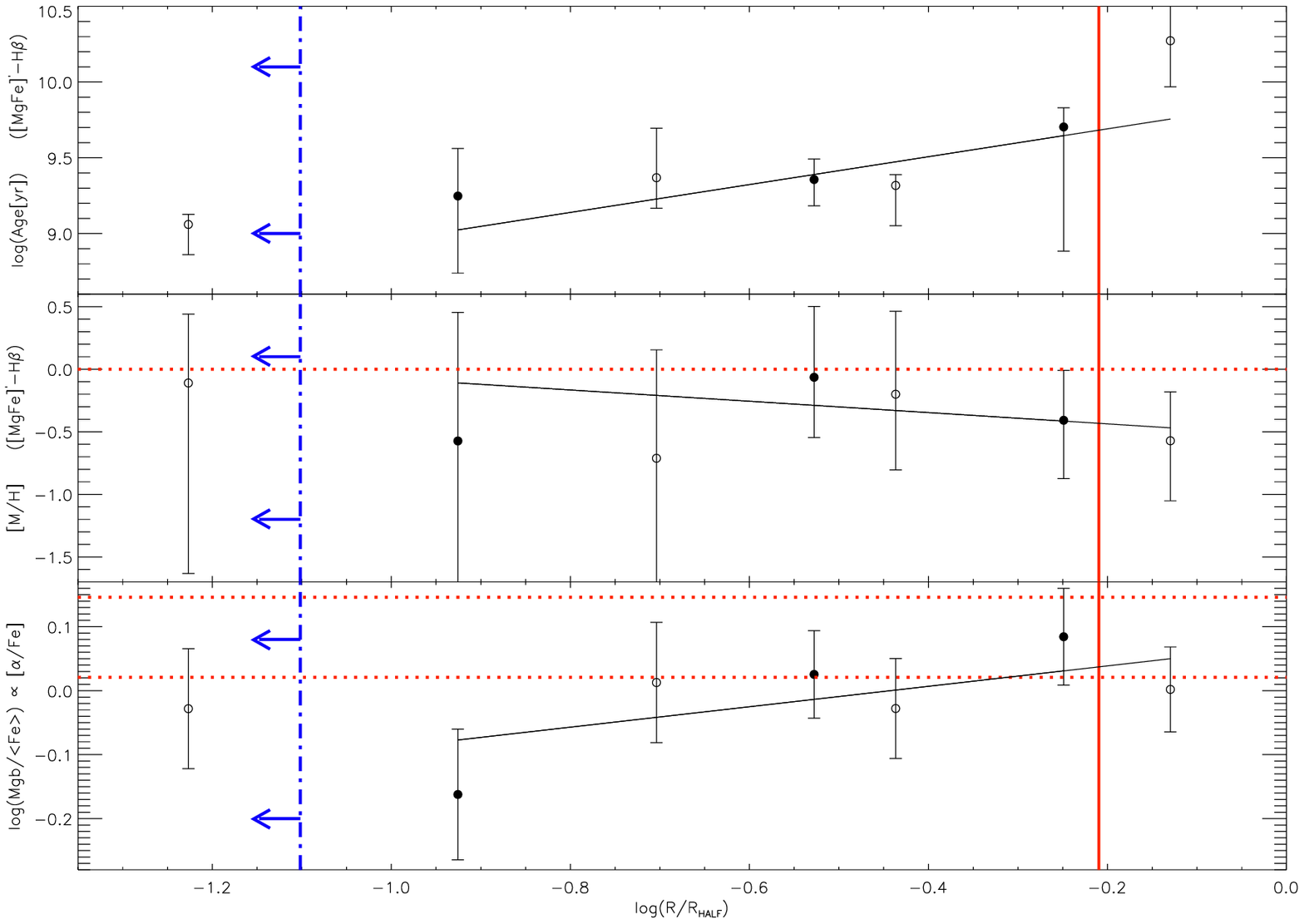}
\end{center}
\caption{\label{fig:PGradE002} Same as Fig.~\ref{fig:PGradN1316} for ESO\,359-G002. Filled (open) symbols correspond to the SW (NE) profile projected in the sky.}
\end{figure*}

\begin{table*}
\begin{center}
\caption{\label{tab:GradAZA} Age, [M/H] and Mg$b$/$\rm \langle Fe\rangle$ Gradients in Fornax S0s}
\begin{tabular}{@{}lc@{\,\,\,}c@{\,\,\,}c@{\,\,\,}c@{\,\,\,}c@{\,\,\,}c@{\,\,\,}c@{\,\,\,}c@{\,\,\,}c@{\,\,\,}c@{\,\,\,}c@{\,\,\,}c@{}}
\hline
\hline
 Name & Grad(log\,Age[yr]) & Grad[M/H] & Grad($\alpha$) \\
   (1)   &      (2) &     (3)  &   (4)   \\
\hline
\hline
\\[0.2em]
  
 NGC\,1316   &   \phantom{1} 0.38$_{-0.13}^{+0.14}$  &  {\bf $-$0.60$^{+0.15}_{-0.10}$}   &   $-$0.02  (0.02)     \\\\
 NGC\,1380  &  $-$0.08$^{+0.09}_{-0.14}$  &  $-$0.21$^{+0.18}_{-0.03}$   &   {\bf $-$0.08  (0.02)}     \\\\
 NGC\,1381   & $-$0.03$^{+0.07}_{-0.10}$  &  $-$0.16$^{+0.10}_{-0.04}$   &   {\bf $-$0.07  (0.01)}     \\\hline
 NGC\,1380A &  \phantom{1} 0.20$^{+0.13}_{-0.08}$  &  $-$0.34$^{+0.14}_{-0.11}$   &     \phantom{1} 0.01  (0.02)     \\\\
 NGC\,1375    &  \phantom{1} 0.46$^{+0.31}_{-0.04}$  &  {\bf$-$0.72$^{+0.12}_{-0.18}$}   &     \phantom{1} {\bf 0.10  (0.03)}     \\\\
 IC\,1963        &  \phantom{1} 0.11$^{+0.09}_{-0.12}$  &  $-$0.26$^{+0.14}_{-0.12}$   &   $-$0.01  (0.02)     \\\hline
 ESO\,358-G059       & \phantom{1} 0.30$^{+0.65}_{-0.64}$  &  $-$0.68$^{+1.45}_{-1.78}$    &  $-$0.11  (0.11)     \\\\
 ESO\,358-G006     &  $-$0.07$^{+0.24}_{-0.20}$  &    \phantom{1} 0.08$^{+0.51}_{-0.51}$    &  $-$0.00  (0.04)     \\\\
 ESO\,359-G002     &   \phantom{1} 0.92$^{+0.41}_{-0.54}$  &  $-$0.45$^{+1.23}_{-0.96}$    &   \phantom{1} 0.16  (0.13)       \\[0.5em]          

\hline
\hline
\end{tabular}\\
\end{center}

\begin{flushleft}
\footnotesize{Notes: For (2) and (3), asymmetric $1\,\sigma$ RMS uncertainties included; For (4), $1\,\sigma$ RMS errors between parentheses; 
Col (2) and (3): SSP Age and [M/H] from H$\beta$ vs.\,[MgFe]' diagram; Col (4): Gradient in log(Mg$b$/$\rm \langle Fe\rangle$) as an $\alpha$-abundance tracer.
Significant gradients ($\ge 3\,\sigma$ detections) are highlighted in {\bf bold}.}
\end{flushleft}
\end{table*}

\bsp

\label{lastpage}

\end{document}